\documentclass[12pt]{article}

\usepackage{graphicx,color}
\usepackage{amssymb}
\usepackage{amsmath}

\newcommand{\be}{\begin{equation}}
\newcommand{\ee}{\end{equation}}
\newcommand{\bea}{\begin{eqnarray}}
\newcommand{\eea}{\end{eqnarray}}
\newcommand{\ba}{\begin{array}}
\newcommand{\ea}{\end{array}}
\newcommand{\eq}[1]{(\ref{#1})}
\def \nn {\nonumber}

\numberwithin{equation}{section}

\begin{document}
\baselineskip=16pt
\begin{titlepage}

\begin{center}
\vspace*{12mm}

{\Large\bf%
Unruh Effect and Holography
}\vspace*{10mm}

Takayuki Hirayama$^{1,2}$\footnote{e-mail:
{\tt hirayama@phys.cts.nthu.edu.tw}},
Pei-Wen Kao$^{2,3}$\footnote{e-mail:
{\tt kao06@math.keio.ac.jp}},
Shoichi Kawamoto$^{2}$\footnote{e-mail:
{\tt kawamoto@ntnu.edu.tw}}
and
Feng-Li Lin$^{2}$\footnote{e-mail:
{\tt linfengli@phy.ntnu.edu.tw}}
\vspace*{5mm}

${}^1$
{\it Physics Division, National Center for Theoretical Sciences, Hsinchu 300, Taiwan
\\[3mm]
${}^2$
Department of Physics, National Taiwan Normal University, Taipei 116, Taiwan
\\[3mm]
${}^3$
Department of Mathematics, Keio University, 3-14-1, Hiyoshi, Kohoku-ku, 223-8522 Yokohama, Japan
}
\\[1mm]

\end{center}
\vspace*{10mm}

\begin{abstract}\noindent%
We study the Unruh effect on the dynamics of quarks and mesons in the context of
AdS$_5$/CFT$_4$ correspondence. We adopt an AdS$_5$ metric with the boundary
Rindler horizon extending into a bulk Rindler-like horizon, which yields
the thermodynamics with Unruh temperature verified by computing the boundary stress
tensor. We then embed in it a probe fundamental string and a $D7$ brane
in such a way that they become the dual of an accelerated quark and a meson in
Minkowski space, respectively. Using the standard procedure of holographic
renormalization, we calculate
the chiral condensate, and also the spectral functions for both the accelerated
quark and meson. Especially, we extract the corresponding strength of
random force of the Langevin dynamics and observe that it can characterize the
phase transition of meson melting. This result raises an issue toward a formulation of 
complementarity principle for the Rindler horizon. We find most of the dynamical
features are qualitatively similar to the ones in the thermal bath
dual to the AdS black hole background, though they could be quite different quantitatively.

\vspace{3ex}


 \end{abstract}

\end{titlepage}

\newpage

\section{Introduction}

A constantly accelerated observer in Minkowski space sees a thermal
radiation at a temperature which is proportional to the value of acceleration.
This is the well-known Unruh effect \cite{Unruh1976} for quantum field theory defined on a
nontrivial geometry. The comoving frame (Rindler frame) of the accelerated observer is the Rindler space
which has an event horizon with the Hawking temperature given above.
Regarding the novelty of the phenomenon, it is natural to ask if one
can observe the Unruh effect in the lab experiments. However, the
direct observation is hard to achieve due to the tiny size of the
Unruh temperature unless the acceleration is very huge. This is the
same situation for observing the Hawking temperature for real black
hole.

Related to the above, we can consider the accelerated charged
particle, and formulate the problem as following. It is well-known
that an accelerated charged particle will radiate as seen in the lab
frame\footnote{Note that this is different from the so called  ``Unruh
  radiation", which is sometimes discussed to be controversial.
See, for example, \cite{Ford:2005sh}.},
and the energy loss rate is described by the Lienard formula as for
classical electrodynamics, and also for non-linear Yang-Mills fields,
which was verified in  \cite{Mikhailov:2003er} by using AdS/CFT
correspondence. Furthermore, the momentum broadening in the
dynamics of an accelerated quark was also observed due to the stochastic nature of the radiation.
See \cite{Dominguez:2008vd} and the references therein.
A natural question arises if we switch to the co-moving frame of the
accelerated quark; how we could interpret the above energy
loss and  momentum broadening in the comoving frame? A naive
expectation based upon the principle of black hole complementarity
\cite{Susskind:1993ki} makes us believe that near the event horizon
the Rindler observer should see something complementary to and
consistent with what the lab observer sees. Moreover, if the
complementarity invokes Unruh effect, then it provides some evidence
for observing Unruh effect as a physical effect indirectly. Indeed,
since the Rindler observer of the accelerated quark sees a thermal
bath of dense quark-gluon plasma, it may lose its energy through the
Brownian motion governed by the Langevin dynamics, and the same
mechanism also for the momentum broadening. This was in fact shown by Xiao in
\cite{Xiao:2008nr}  by using the AdS/CFT correspondence to extract the strength of the random forces in Langevin dynamics, and the resultant momentum broadening agrees with the one from the stochastic radiation picture based on  \cite{Dominguez:2008vd}.

However, the above complementarity should be placed under more
scrutiny. Note that the above setting is in the strongly coupled
regime with the help of AdS/CFT correspondence, it is not clear if the
complementarity holds also for the weak coupling regime.  Naively, it
does not work for accelerated neutral elementary particle (zero
coupling) since there is no radiation from it in the lab frame, but
the Rindler observer still sees the thermal bath and thus the random
kicks causing the Brownian motion. This casts doubt on the
complementarity for the Rindler horizon, and deserves further
study. Also, how about the complementarity for an accelerated neutral
composite particle such as meson?


  Motivated by the above, in this paper we discuss the quark and meson dynamics caused by the
Unruh effect in ${\cal N}=4$ Supersymmetric Yang-Mills theory (SYM)
from the holographic description\footnote{Some related works considering the Unruh
effect in AdS space based on  Global Embedding in Minkowskian Space-time method
(the GEMS \cite{Deser:1998xb}) were done in \cite{Paredes:2008cr}.}. Though we are not trying to
resolve the above issues, our results on the accelerated meson in comoving frame will help to shed some light
on complementarity regarding the Rindler horizon invoking the Unruh effect.

We start with the metric seen by the comoving observer of an accelerated
string in the AdS$_5$ space found in \cite{Xiao:2008nr}.
As usual, the radial direction in the above geometry is identified with the energy scale of
boundary CFT which is in the comoving frame of an accelerated observer in
Minkowski space. We then calculate its holographic stress tensor, and verify that it
is that of a thermal vacuum with nonzero temperature, entropy and energy density.
Moreover, the temperature agrees with the Unruh temperature and the first law of
thermodynamics is satisfied.  This is surprising because the bulk spacetime seen by the comoving observer of the accelerated string is nothing but the pure AdS space up to a coordinate change.  It is then curious to see the differences between the our thermal vacuum and the one associated with the AdS-Schwarzschild space when considering the dynamical properties.

The Rindler space covers only a part of Minkowski space. Therefore, the pure vacuum
state of the Minkowski spacetime restricted to one of the Rindler wedges becomes a
mixed thermal state in Rindler spacetime because the states corresponding to the other
Rindler wedge are traced out.
Correspondingly, the comoving observer of the accelerated string also only sees a part of the AdS$_5$ space and the horizon of the accelerated observer is just the bulk extension of the boundary Rindler horizon.
We then expect that the boundary Rindler thermal property will be holographically dual to the  thermodynamical properties associated with the bulk horizon. This is indeed the case as can be seen from the holographic stress tensor.
Furthermore, we introduce a probe D7 brane which is static in the comoving frame. Then the scalar field on this D7 brane is dual to the operator $\bar{q}q$ or meson field which is accelerated in Minkowski space. We then compute the value of the chiral condensate which properly shows the behavior in finite temperature\footnote{%
If we introduce a D7 brane in a supersymmetric way \cite{Karch:2002sh} which is dual to
$\bar{q}q$ defined in the Minkowski
space, move to the comoving frame without changing the way of the D7 brane embedding, and compute the value of the chiral condensate, we do not obtain the behavior of finite temperature. This is expected since the expectation value does not depend on the coordinate frame~\cite{Unruh:1983ac}.}.  Other dynamical properties such as the quark-anti-quark potential and the meson spectral function related to the Langevin dynamics will also be evaluated in this context.

This paper is organized as follows.
In section 2, we prepare a constantly accelerated open string in AdS$_5$ space and
define a comoving frame where the string is static.
This is then dual to a quark which is constantly accelerated in Minkowski spacetime.
Identifying the radial direction in the comoving frame as the energy scale in the CFT
side, we holographically compute the stress energy tensor
and verify its thermal nature by the expected thermodynamical relation. Then
we calculate the quark and anti-quark potential and the spectral functions of the
constantly accelerated quark. In section 3, We introduce a probe
D7 brane on which a constantly accelerated string ends perpendicularly.
This is thus dual to the meson
which is constantly accelerated in Minkowski spacetime.
We then holographically compute the value of the chiral condensate, and its
spectral functions. We find that all the above quantities
bear the similar qualitative feature to their thermal counterparts which are dual to the fundamental string and D7 brane in the AdS black hole background.
From the spectral functions, we extract the corresponding strength
of random force of the Langevin dynamics,
and observe that there happens a significant change of its strength
around the transition point suggested by the analysis of the one-point function.
We also provide an interpretation for this based on a meson dissociation
picture. In section 4 we conclude our paper with some discussions and open questions. In Appendix A, we discuss an issue about the choice of the D7 brane embedding. In
Appendix B, we provide the details for deriving the flow equation of the
retarded Green function of the accelerated meson.

\section{Holography for an accelerating observer}

In this section, we first review the accelerated string solution in AdS$_5$
space~\cite{Xiao:2008nr}, this is dual to an accelerated quark in the boundary dual
SYM. We then transform to the comoving frame of the accelerated string and
consider the thermodynamics of the dual SYM in Rindler space via AdS/CFT correspondence. We also evaluate the holographic retarded Green function of the
accelerating quark and numerically extract its spectral function.
We recognize that the feature of spectral function is
qualitatively similar to but quantitatively
different from the one for the usual thermal vacuum dual to the
bulk black hole. Let us start with the metric of AdS$_5$ space given by
\begin{align}
 ds_{AdS_5}^2 &= \frac{R^2}{z^2} \left(
  -dt^2 + dx_1^2 + dx_2^2 + dx_3^2 + dz^2
  \right)
  ,
 \label{adsp}
\end{align}
where $R$ is the AdS curvature. We consider a fundamental string which is
accelerating along $x_1$ direction whose trajectory is given by
\begin{align}\label{accF1}
 x_1^2 - t^2 = a^{-2} - z^2 ,
\end{align}
where $a$ is the value of acceleration.
This solution satisfies the equations of motion from Nambu-Goto action~\cite{Xiao:2008nr}.
We then define a coordinate transformation such that this fundamental string becomes static in the new coordinate system.
The new coordinates are defined as follows:
\begin{align}
 x_1 &= \sqrt{a^{-2}-r^2} e^{\alpha a} \cosh (a \tau) ,
 \hspace{3ex}
 t = \sqrt{a^{-2}-r^2} e^{\alpha a} \sinh (a \tau) ,
 \hspace{3ex}
 z = r e^{\alpha a},
\end{align}
and the equation of motion \eq{accF1} of the accelerated string becomes
$\alpha=0$,  i.e., the string is static in the new coordinate
system. This coordinate system is Rindler-like and only covers the right wedge of
Rindler space $0<x_1<|t|$ for fixed $z$.
See Fig.~1.
\begin{figure}[t]
 \begin{center}
\vspace{-2.5em}
  \includegraphics[height=9cm]{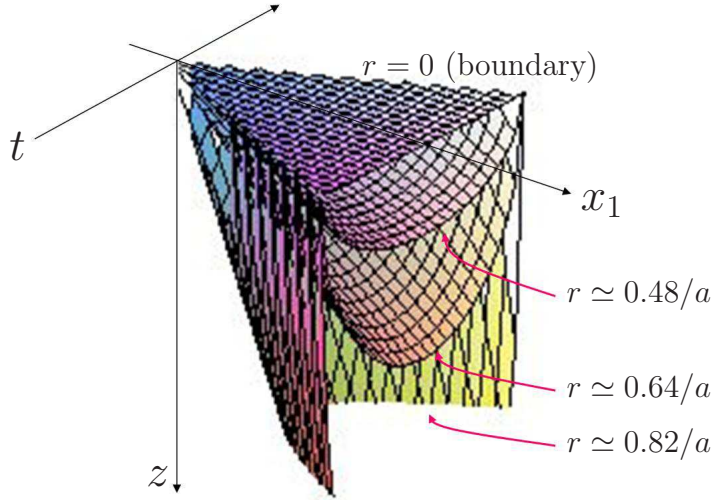}
 \put(-233,170){\makebox(0,0){\Large $t$}}
 \put(-12,148){\makebox(0,0){\Large $x_1$}}
 \put(-180,45){\makebox(0,0){\Large $z$}}
 \put(0,114){\makebox(0,0){ $r \simeq 0.48/a$}}
 \put(0,78){\makebox(0,0){ $r \simeq 0.64/a$}}
 \put(0,57){\makebox(0,0){ $r \simeq 0.82/a $}}
 \put(-60,200){\makebox(0,0){ $r=0$ (boundary)}}
\vspace{-8ex}
\caption{The constant-$r$ surfaces in $(x_1,t,z)$ coordinates. The
  comoving frame covers only the right Rindler wedge. As $r$ gets
  larger, constant-$r$ surface gets steeper. The AdS boundary is at
  $r=0$, and the bulk extension of the boundary Rindler horizon is
  located at $r=1/a$.}
 \end{center}
 \label{r_surface}
\end{figure}
The metric of AdS$_5$ in the new coordinates is written as
\begin{align}
 ds_{AdS_5}^2 &= \frac{R^2}{r^2}\Big[
 \frac{dr^2}{1-a^2r^2} - (1-a^2r^2) d\tau^2 +d\alpha^2
 + e^{-2 \alpha a} (dx_2^2 + dx_3^2) \Big]
 .
 \label{29}
\end{align}
This metric is also given in \cite{Xiao:2008nr}.
Note that the boundary metric is conformal to Rindler space,
\begin{align}
 ds_{bdry}^2 &=  - d\tau^2 +d\alpha^2
 + e^{-2 \alpha a}  (dx_2^2 + dx_3^2)=e^{-2\alpha a}[e^{2\alpha a} (-d\tau^2+d\alpha^2)+dx_2^2+dx_3^2]\;.
 \label{bdm}
\end{align}
Obviously, the metric (\ref{29}) has a horizon at $r=1/a$, i.e., the bulk extension of the boundary Rindler  horizon at $t=\pm x_1$, and the associated Hawking temperature is $T=a/2\pi$ which is the same as the Unruh temperature at the
boundary. Therefore, the temperature in the CFT side is the same as that in
the bulk spacetime.  Thus we identify the new radial coordinate $r$ with
the energy scale of dual CFT in Rindler space.

We shall remind that the metric \eq{29} does not
describe a black hole in AdS space since it is just obtained by a
coordinate transformation from the AdS metric \eq{adsp} of Poincare
coordinates. However, since we identify the $r$ coordinate
which is different from $z$ coordinate in the metric \eq{adsp} as the energy
scale in CFT, the CFT dual is defined in the boundary Rindler metric~\eq{bdm}
but not in the Minkowski one. This coordinate transformation induces
a horizon, and thus leads to thermodynamics of the dual SYM as
shown below. Nevertheless, we will see that the holographic spectral function of
the accelerated quark from the metric (\ref{29}) is different from the one of
thermal quark heated by the bulk AdS black hole.
For later use, we further define a new radial coordinate $w$ as
\begin{align}
 r = \frac{4w}{a^2w^2+4} \,,
\end{align}
and the AdS$_5\times S^5$ metric becomes
\begin{align}
 ds_{AdS_5\times  S^5}^2
 &= \frac{w^2}{R^2}
 \Big[
  - (1-\frac{a^2R^4}{4w^2})^2d\tau^2
  +(1+\frac{a^2R^4}{4w^2})^2
   ( d\alpha^2+ e^{-2\alpha a} (dx_2^2 + dx_3^2)) \Big]
 \nonumber
 \\
 &\;\;\;
 +R^2\frac{dw^2}{w^2}
 +R^2 ( d\psi^2 + \cos^2\psi d\theta^2
 +\sin^2\psi d\Omega_3^2)
 .
\end{align}
The metric in the second line is just a 6-dimensional flat space,
and for it we introduce the 4-dimensional polar coordinates
($\rho$ and $\Omega_3$) and the 2-dimensional flat coordinates
($w_5$ and $w_6$). In terms of these coordinates, the metric becomes
\begin{align}
 ds_{AdS_5 \times S^5}^2
 &= \frac{w^2}{R^2}
 \Big[
  - (1-\frac{a^2R^4}{4w^2})^2d\tau^2
  +(1+\frac{a^2R^4}{4w^2})^2
   ( d\alpha^2+ e^{-2\alpha a} (dx_2^2 + dx_3^2)) \Big]
 \nonumber
 \\ \label{xiao3}
 &\;\;\;\;
 +\frac{R^2}{w^2} ( d\rho^2 + \rho^2 d\Omega_3^2 + dw_5^2 + dw_6^2 )
,
\end{align}
where $ w^2=\rho^2 + w_5^2+w_6^2$.
The AdS boundary and the horizon are located at $w=\infty$ and
$w=aR^2/2$, respectively.

\subsection{Thermodynamics}

From the previous discussions, the metric \eq{29} describes the
holographic dual gravitational background for CFT defined in Rindler metric~\eq{bdm},
and the coordinate $r$ is identified as the energy
scale of CFT. We may expect the Unruh effect generating a thermal
system with the temperature $a/2\pi$. Note that the thermal setting is
quite different from holographic dual of the AdS black hole.

To see this, we compute the stress energy tensor of CFT from the bulk metric using
AdS/CFT dictionary \cite{Skenderis:2002wp}. The bulk action is given by
the standard Einstein action plus the cosmological constant and the
boundary action,
\begin{align}
 S &=
 -\frac{1}{16\pi G_5} \!\int_{\cal M}\! d^{5}x \Big[ R -2\Lambda \Big]
 -\frac{1}{8 \pi G_5} \!\int_{\cal \partial M}\! d^4x \sqrt{-\gamma} \;
 \Theta,
\\
 &\;\;\;\;
 \gamma_{\mu\nu} = g_{ab} - n_an_b
 , \hspace{3ex}
 \Theta^{\mu\nu} = - ( \nabla^\mu n^\nu + \nabla^\nu n^\mu )
 ,
\end{align}
where $n_{\mu}$ is the unit vector normal to the boundary ${\cal \partial M}$.
Then we substitute the background metric~\eqref{29} into the metric of
the action and compute its on-shell value. Since the value is divergent,
we introduce a cutoff near the AdS boundary $r=\epsilon$, i.e., ${\cal
\partial M}$ is the $r=\epsilon$ surface, and add the counter terms
defined at $r=\epsilon$. In our case the counter term action  takes the form as
\begin{align}
 S_C &= \frac{1}{8 \pi G_5} \!\int_{\cal \partial M}\! d^4x
 \sqrt{-\gamma} \left[ c_1 + c_2 R_{\gamma} \right]
 ,
\end{align}
where $R_{\gamma}$ is the Ricci scalar constructed from the induced
metric $\gamma$, and $c_1$ and $c_2$ are determined so that the total
action $S+S_C$ is finite.
Thus, we find that $c_1=6/R$ and $c_2=-R^3$.
According to the AdS/CFT correspondence, the stress
energy tensor of CFT is computed as follows,
\begin{align}
 T^{\mu}_{\nu} &= \lim_{\epsilon\rightarrow 0}
  \frac{1}{\sqrt{-\gamma}}\frac{\delta (S+S_C)}{\delta \gamma_{\mu}^{\nu}}
 =
 \lim_{\epsilon\rightarrow 0} \frac{1}{8\pi G_5} \Big[
 {\Theta}^{\mu}_{\nu} -{\Theta} \gamma^{\mu}_{\nu}
 +\frac{1}{\sqrt{-\gamma}}
 \frac{\delta S_{C}}{\delta \gamma_{\mu}^{\nu}} \Big] \
 \nonumber
 \\
 &=
 \frac{1}{8\pi G_5}
 \left[
 \frac{a^4 R^3}{4}\; \mbox{diag} (3,-1,-1,-1)
 \right] \;.
\end{align}

This stress tensor appears as that of conformal thermal gas
at the temperature $T=a/2 \pi$, i.e.,  the energy  density
${\cal E}$ and the pressure $p$ satisfy the relation $ {\cal E}= 3 p \propto T^4$,
the equation of state for conformal gas. Moreover,
the entropy density of the dual SYM can be read off from the area of the horizon, which is
$s={1\over 4 G_5} a^3 R^3$\footnote{
The horizon area is given $A=R^3a^3\!\int\! d\alpha dx_2dx_3$, and then
the entropy density is given by $A/( 4G_5\! \int \! d\alpha dx_2 dx_3\sqrt{-\gamma} )$.
}, and the thermal
quantities satisfy the first law of thermodynamics, i.e., $p+{\cal E}=Ts$.
This shows that the system is in a thermal equilibrium as expected from the Unruh effect.

\subsection{Quark-anti-quark potential for accelerated quarks}

We also compute the quark and anti-quark potential in Rindler space from the
Wilson loops and see whether the potential shows an appropriate behavior
at a finite temperature. The holographic counterpart of the
Wilson loop is the worldsheet of a fundamental string with endpoints being identified as a quark and anti-quark pair~\cite{Maldacena:1998im}. We evaluate
the worldsheet profile in the metric (\ref{29}) to extract the potential between
quark and anti-quark, both of which are constantly accelerated along the same direction.

For simplicity, we separate the quark and anti-quark only along $\alpha$
direction in the Euclidean version of the metric (\ref{29}) and thus we study an open string
whose profile is given by $\alpha=\alpha(r)$. This profile describes a
string extending into the bulk but with its end points fixed on the
boundary. The Euclidean Nambu-Goto action and the equation of motion from that are
\begin{align}
 & S = T_F\!\int\! d\tau dr \frac{R^2}{r^2}
 \sqrt{h(r)(\partial_r\alpha)^2 +1}
 ,
 \\
 & \frac{R^2}{r^2}
 \frac{h(r)\partial_r\alpha}{\sqrt{h(r)(\partial_r\alpha)^2 +1}}
 =
 \frac{R^2}{r_0^2}\sqrt{h(r_0)}
 ,
 \end{align}
respectively. Here  $T_F$ is the string tension
(and we will take the normalization $T_F=1$),
$h(r)=1-a^2 r^2$, and we require the boundary condition at the turning point
of the string profile $r=r_0$ such that
$\partial_r \alpha|_{r=r_0}=\infty$. Then the distance between the quark
and anti-quark $L$ and the total energy $E$ are given by
\begin{align}
 L &= 2 \!\int^{r_0}_{\epsilon}\! \frac{dr}{\sqrt{
 \frac{r_0^4}{r^4}\frac{h(r)^2}{h(r_0)} - h(r) } }
 ,
 \\
 E &= 2 \!\int^{r_0}_{\epsilon}\! dr \frac{R^2}{r^2\sqrt{1-
 \frac{r^4h(r_0)}{r_0^4h(r)}}},
\end{align}
respectively. Here  $\epsilon (\approx 0)$ is the UV cutoff. On the other hand,  the
mass of the free quarks, defined by the energy of two parallel straight
strings configuration extending from $r=\epsilon$ to the horizon at
$r=1/a$, is given by
\begin{align}
 M &= 2 \!\int^{1/a}_{\epsilon}\! dr
 \frac{R^2}{r^2}=
 2R^{2} \left(
\int_{\epsilon}^{r_{0}}dr \ \frac{1}{r^{2}}+\frac{1}{r_{0}}-a
\right) \,.
\end{align}

Thus, the quark-anti-quark potential $V_{\bar{q}q}$ is given by
\be
V_{\bar{q}q}(L)=E-M.
\ee

\begin{figure}[t]
 \begin{center}
  \includegraphics[width=17em,clip]{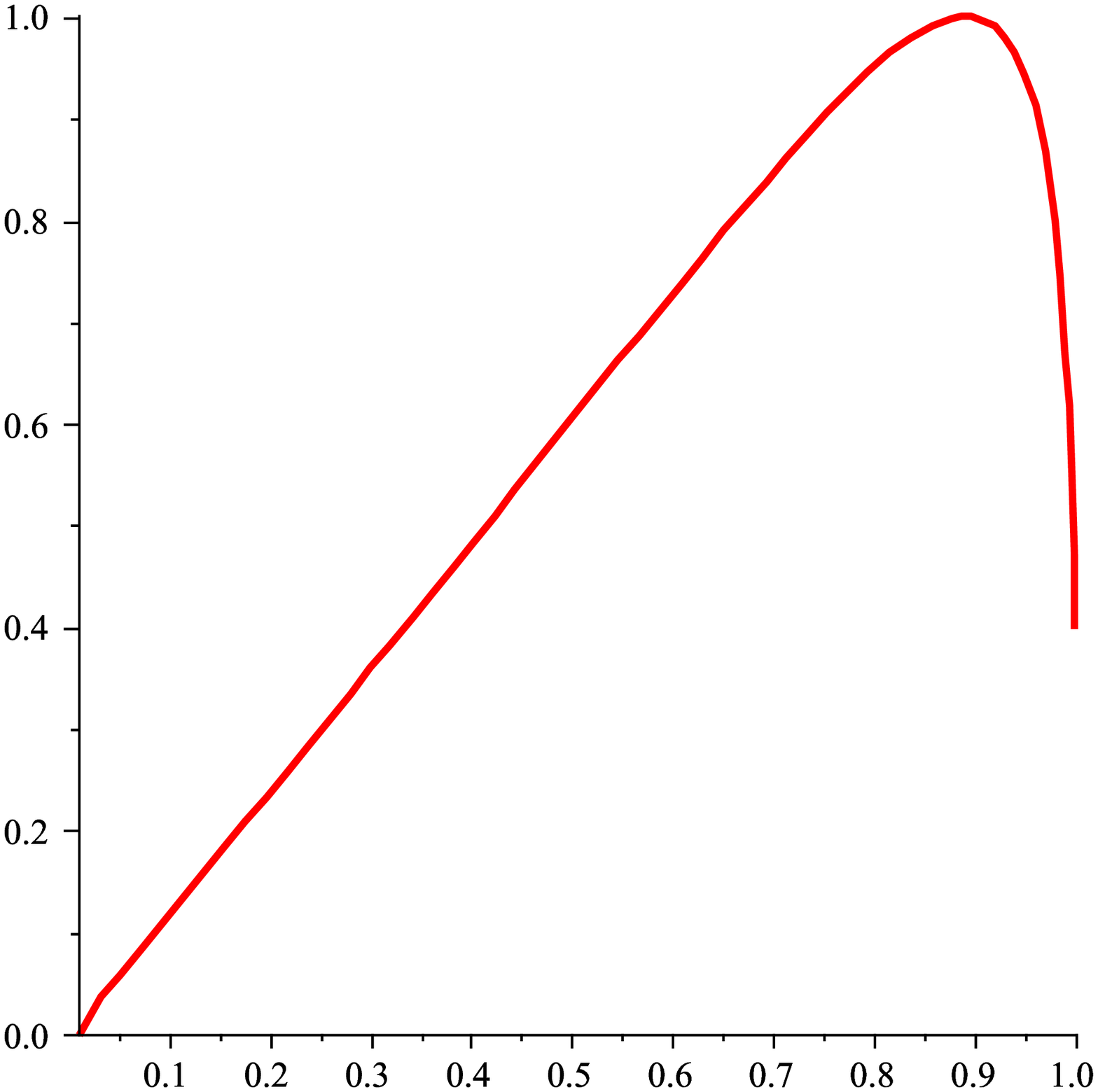}
\hspace{2em}
  \includegraphics[width=17em,clip]{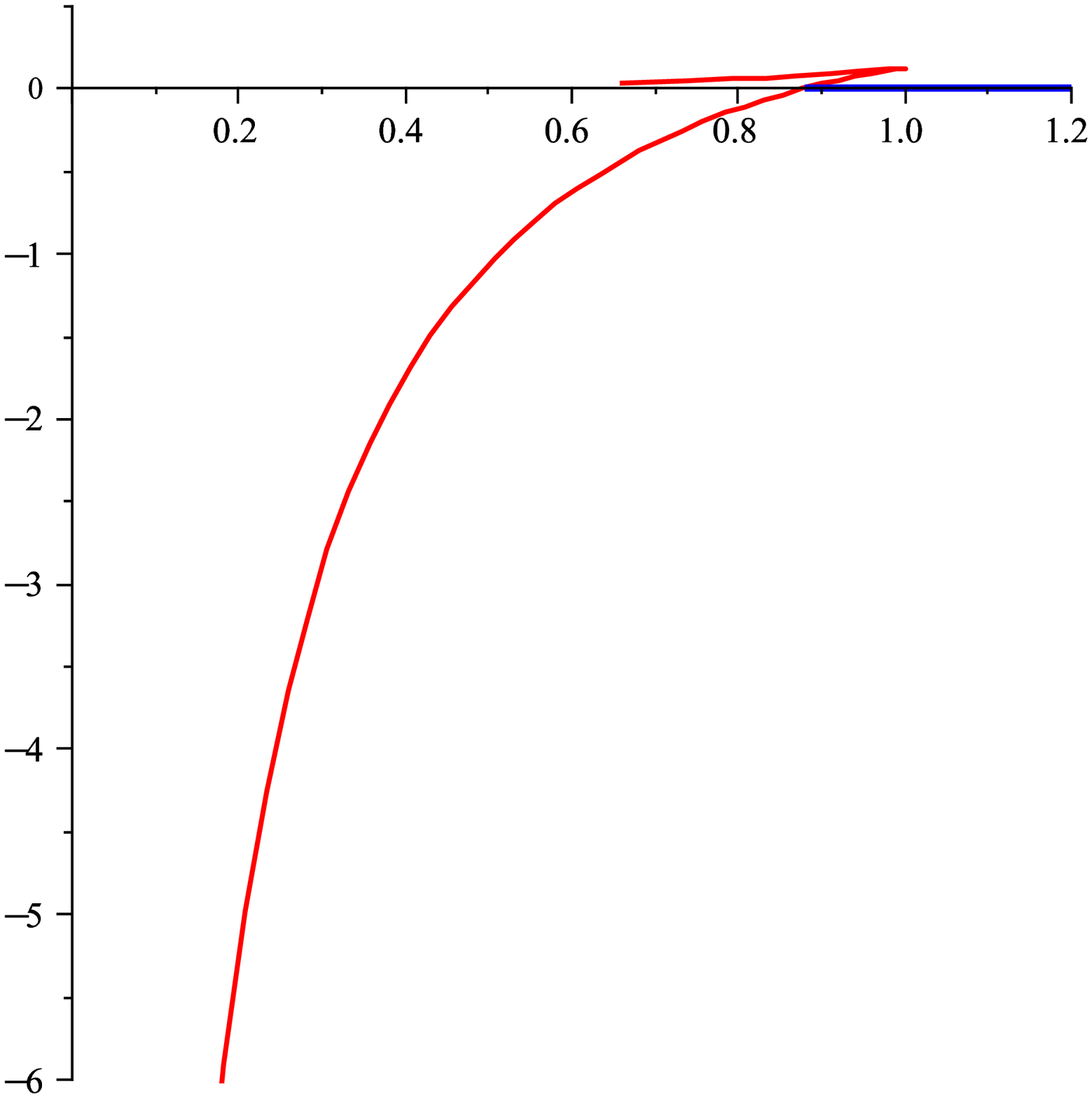}
\put(-420,205){\makebox(0,0){ $L$}}
\put(-325,-5){\makebox(0,0){ $r_0$}}
\put(-197,195){\makebox(0,0){ $V_{\bar{q}q}$}}
\put(-10,190){\makebox(0,0){ $L$}}
\put(-39,165){\makebox(0,0){ $\uparrow$}}
\put(-53,165){\makebox(0,0){$\uparrow$}}
\put(-30,152){\makebox(0,0){ $L_{max}$}}
\put(-52,152){\makebox(0,0){$L_c$}}
 \end{center}
 \caption{The quark-anti quark separation and the potential.
All the plots are with $a=1$ and $R=1$.
{\bf Left}: $L$ against $r_0$.
As $r_0$ gets larger,
the separation becomes larger until $r_0=0.89/a$ and starts to decrease for $r_0>0.89/a$.
Thus, there is  a maximal value for $L$, $L_{max}=1.0/a$.
{\bf Right}:
The potential $V_{\bar{q}q}$ against $L$.
At $L=L_c \sim 0.88/a < L_{max}$, the potential vanishes. }
 \label{fig_qq_plot}
\end{figure}

After some numerical work in evaluating the quark-anti-quark potential,
we find the Coulomb potential behavior for small $L$ region compared to $1/a$.
The numerical fitting gives $(R=1)$
\be
V_{\bar{q}q}(L) \simeq  -\frac{1.43}{L}+1.97 \ a \,,
\ee
where the numerical values are
chosen to fit the values in $L \leq 0.1/a$ regime.
This Coulomb behavior is expected for ${\cal N}=4$ SYM at zero temperature
and its coefficient is very close to the zero-temperature result $4T_F R^2 \Gamma(3/4)^4/2\pi$~\cite{Maldacena:1998im},
where the small deviation is due to the fitting by using only finite number of terms.
The second term describes the screening effect which reduces the binding energy, however, its behavior is quite different from the one in the AdS-Schwarzschild background, which is proportional to $L^3$.

As we separate the quark and anti-quark further by increasing $r_0$,
we find the binding energy becomes zero around $L=L_c \sim 0.88/a$
(with $r_0\sim 0.73/a$). If we simply increase $r_0$ further,
$L$ becomes larger, takes a maximal value $L_{max} \sim 1.0/a$ with
$r_0\sim 0.89/a$ and starts decreasing until $r_0=1/a$.
The binding energy is always
negative, i.e. $V_{\bar{q}q}>0$ for $r_0 > 0.73/a$.
The numerical results of the quark-anti quark separation and the potential are shown
in Fig.~\ref{fig_qq_plot}.
Since the binding energy is
negative for  $L>L_c$, the configuration of two straight strings has lower energy than that of a single connected string, and thus becomes the dominant one in this region. That is, the quark and anti-quark no longer make a bound state for $L>L_c$ and start moving freely. These behavior is similar to those computed from AdS black hole \cite{WL_finiteT}.

   In the lab frame, the screening is due to the squeezed lightcone caused by the  acceleration so that the causal gauge interaction will be effected in a shorter range.  Turning to the comoving frame, it is then interpreted as the thermal effect.  However, from the bulk point of view, the real thermal effect and the Unruh one correspond to different bulk geometries with event horizons, causing different squeezing of the lightcone and thus different screening.  This suggests that we indeed can tell the Unruh effect from the real thermal effect (at least quantitatively) by looking into the dynamical properties such as $V_{\bar{q}q}$, we see more examples later on. Moreover, once part of the open string connecting the quarks sinks behind the horizon, it loses the causal contact with the part outside the horizon, and can be understood the energy loss due to radiation in the lab frame, or due to random kicks in the comoving frame. Therefore, the energy loss of the accelerated or thermal quark has an unified geometric picture in the bulk AdS.


\subsection{Spectral function of an accelerated quark}

\label{sec:spec_string}

A quark in the thermal background is subjected to the random force of
the Langevin dynamics dictated by the fluctuation-dissipation theorem.
The Langevin equation is
\be
{d p_i \over d\tau}=-\gamma p_i +{\cal F}_i(\tau)
\ee
where $\gamma$ is the friction coefficient, and the thermal kick is
characterized by the random force ${\cal F}_i$ with the following average:
\be
\langle {\cal F}_i (\tau) \rangle=0\;, \qquad
\langle {\cal F}_i(\tau) {\cal F}_j(\tau') \rangle
=\kappa \delta_{ij} \delta(\tau-\tau')\;.
\ee
The coefficient $\kappa$ is the strength of the random force.
Furthermore,  by the Kubo formula one can extract the strength of the
random force from the retarded Green function.

In \cite{Xiao:2008nr} the author modeled the Brownian motion of the
accelerated quark by the Langevin dynamics, and then analyzed the low energy
behavior of the holographic retarded Green function of the accelerated quark,
from which the author found that the strength of the random force $\kappa$ is
\be\label{xiaokappa}
\kappa=\lim_{\omega \rightarrow 0} {-2T\over \omega}
\text{Im}\; G_R(\omega)={\sqrt{\lambda} \over 2\pi^2} a^3
\ee
where $\lambda$ is the 't Hooft coupling constant, and the $\sqrt{\lambda}$ dependence is the
generic feature predicted by the AdS/CFT correspondence for the strong coupling regime. This has the same
coupling and temperature dependence as the one \cite{Giecold:2009cg} for
the quark in the thermal vacuum dual to a string in the AdS-Schwarzschild background.

Despite of having the same form of random forces for the thermal and
the accelerated quarks, we wonder if their spectral functions also show
the same behavior or not. This question arises since
the metric (\ref{29}) is just a coordinate transformation of the Poincare patch of AdS
space, and contains no real black hole. On the other hand, it is known
that \cite{quasinormal} the resonance excitations of the thermal CFT are
dual to the quasi-normal modes of AdS black hole. Therefore, it is not
obvious if the metric (\ref{29}) also admit the quasi-normal modes or
not. If this is the case, then it suggests that we can not distinguish
the Rindler vacuum from the thermal one even by comparing the spectral
functions of accelerated and thermal quarks or mesons.

 Furthermore, the stochastic Brownian motion is characterized by the $p_{\perp}$ and $p_L$ broadening
 \be\label{p-broadening}
 {d p^2_{\perp} \over dt}=2\kappa, \qquad  {d p^2_L \over dt}=\kappa
 \ee
which can then be related to the momentum broadening in the lab frame
by restoring the boost factor $\gamma=\sqrt{a^2t^2+1}$, i.e.,  ${d
  p^2_{(lab)\perp} \over dt}=2\kappa/\gamma$ and   ${d p^2_{(lab)L}
  \over dt}=\kappa \gamma$. The momentum broadening in the lab frame
is due to the stochastic nature of the classical radiation, which can
be obtained by considering the dynamics of the trailing string in pure
AdS space \cite{CasalderreySolana:2006rq}.
Since 't Hooft coupling constant $\sqrt{\lambda}$ can be interpreted
as the parton emission probability in this picture, it seems natural
that $\kappa$ is proportional to this factor.
 In \cite{Xiao:2008nr} it was shown that  the momentum broadening
calculated in the Rindler (using AdS/CFT correspondence) and lab (based on partonic picture with the probability of emitting a gluon being $\sqrt{\lambda}$) frames agree at the large time limit. The result could be
implied by the principle of complementarity for the Rindler horizon as
discussed in the introduction.

  Furthermore, we can consider the spectral function
for an accelerated quark, and then for an accelerated meson in the next section.
In order to compare with the D7 brane (i.e. meson) case, we embed the
fundamental string dual to an accelerated quark into the metric in the form of
\eq{xiao3},
\footnote{We also calculate the spectral function for the metric
which is given by (28) in \cite{Xiao:2008nr} and we find that the qualitative behaviors for
both choices are the same.}
and choose the static gauge
\begin{align}
  (\sigma^0, \sigma^1)
 =& (\tau,\rho) \,.
\end{align}
The embedding ansatz is given by
\begin{align}
  X^\mu =& \left(\tau, \rho, \alpha(\rho,\tau), 0 , \cdots , 0 \right)
 \,.
\end{align}
The Nambu-Goto action in the case becomes
\begin{align}
  S=&
 -T_F \int d\tau d\rho \, \mathcal{L}
 \\
 \mathcal{L}=&
 \sqrt{\left(h_-^2-h_+^2 \dot\alpha^2 \right)
 \left(1+\frac{\rho^4}{R^4}h_+^2 \alpha^{\prime 2} \right)} \,,
\end{align}
where $h_\pm = 1 \pm \frac{a^2 R^4}{4\rho^2}$.
The string profile solution found by Xiao is given by $\alpha=0$, around which
we now consider the fluctuation given by $\alpha(\tau,\rho)$.
By expanding the action with respect to the fluctuation $\alpha$,
we obtain the quadratic order action given by
\begin{align}
 \mathcal{L}_2 =&
 \frac{1}{2}
 \left[
 Q_1 (\partial_\rho \alpha)^2 + Q_2 (\partial_\tau \alpha)^2
 \right]
\end{align}
where
\begin{align}
 Q_1 =& \frac{\rho^4}{R^4}h_- h_+^2 \,,
 \qquad
 Q_2 = -\frac{h_+^2}{h_-} \,.
\end{align}

It is convenient to introduce the dimensionless coordinates
\begin{align}
 \bar\rho=& \frac{1}{aR^2}\rho \,,
 \qquad
 \bar\tau = a\tau \,,
\end{align}
and thus the quadratic action becomes
\begin{align}
 \label{ac_f2_0}
 S_2 =&
 - a^2 R^2 T_F  \int d\bar\tau d\bar\rho \;
 \frac{1}{2}
 \left[
 \bar{Q}_1 (\partial_{\bar\rho} \alpha)^2 + \bar{Q}_2
 (\partial_{\bar\tau} \alpha)^2 \right]
\end{align}
where
\begin{align}
 \bar{Q}_1 =& \bar\rho^4 \bar{h}_- \bar{h}_+^2 \,,
 \qquad
 \bar{Q}_2 = -\frac{\bar{h}_+^2}{\bar{h}_-} \,,
 \qquad
 \bar{h}_\pm = 1 \pm \frac{1}{4 \bar\rho^4} \,.
\end{align}
Note that the horizon is now at $\bar{\rho}=1/4:=\bar{\rho}_h$.

Introducing the Fourier transform of the field
\begin{align}
 \label{fourier_alpha}
 \alpha (\bar\tau,\bar\rho)
 =& \int \frac{d\bar\omega}{2\pi}
 e^{-i\bar\omega \bar\tau} z_{\bar\omega}(\bar\rho) \,,
\end{align}
then from (\ref{ac_f2_0}) the equation of motion is
\begin{align}
 \label{eom_rhoF}
 \partial_{\bar\rho} (\bar{Q}_1 \partial_{\bar\rho} z_{\bar\omega}) - \bar\omega^2
 \bar{Q}_2 z_{\bar\omega} =0 \,.
\end{align}

Since we are interested in the retarded Green function, we can
follow \cite{Iqbal:2008by} to solve the first order flow equation
of the ``running'' retarded Green function directly without first
solving the equation of motion as in \cite{Son:2002sd}. To
do that, we introduce the conjugate momentum,
\begin{align}
 \tilde\pi_\omega = -a^2 R^2 T_F \bar{Q}_1 \partial_{\bar\rho}  z_{\bar\omega}(\bar\rho)
 = -a^2 R^2 T_F \tilde\pi_\omega^0
 \,,
\end{align}
and by the analog of linear response theory \cite{Iqbal:2008by}, the
retarded Green function of the accelerated quark is given by
\begin{align}
 \label{def_Gr}
 G_R(\bar\omega) =&
 -a^2 R^2 T_F
 \lim_{\bar\rho \rightarrow \infty} \xi(\bar\rho) \,,
\end{align}
where we define
\begin{align}
 \label{def_xi0}
 \xi =& \frac{\tilde\pi_{\bar\omega}^0(\bar\rho)}{
 z_{\bar\omega}(\bar\rho)} \,.
\end{align}
Differentiating both sides of (\ref{def_xi0}) with respect to $\bar\rho$
and using the equation of motion, we obtain the flow equation
\begin{align}
 \label{flow_eq}
 \partial_{\bar\rho} \xi =&
 -\frac{1}{\bar{Q}_1} \xi^2
 +\bar\omega^2 \bar{Q}_2 \,.
\end{align}

\begin{figure}[tbh]
 \begin{center}
  \includegraphics[width=17em,clip]{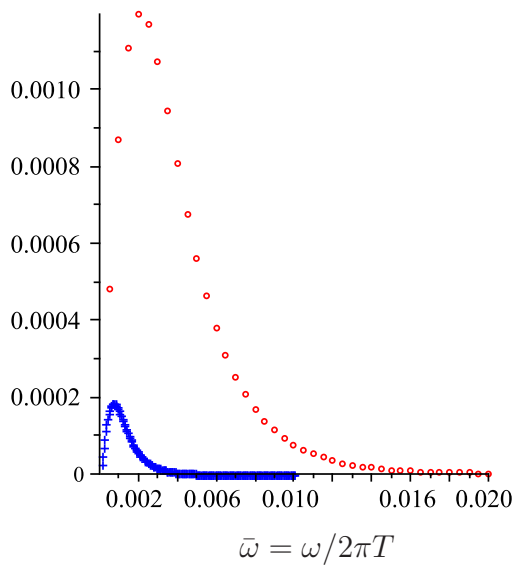}
 \put(-80,-10){\makebox(0,0){$\bar\omega=\omega/2\pi T$}}
 \end{center}
\vspace*{-0.6em}
 \caption{Spectral functions  of  thermal and accelerated quarks against the
 dimensionless frequency $\bar\omega=\omega/2\pi T$.
 The temperature is tuned to be $T=1/2\pi$.
 Red (circle) points for the accelerated quarks and Blue (cross) ones for the thermal quarks.}
 \label{spec_String}
\end{figure}

One can also solve the equation of motion \eq{eom_rhoF} near the AdS
boundary, and obtain the asymptotic form of both $z_{\bar\omega}$ and
$\xi$ as follows,
\begin{align}
 z_{\bar\omega} \sim & \;
 \bar{m} \left( 1 + \frac{\bar\omega^2}{2}\frac{1}{\bar\rho^2} +
 \mathcal{O}(\bar\rho^{-4}) \right)
 +\frac{\bar\nu}{\bar\rho^3}
 \left( 1
 -\frac{1}{20} \left( 2\,{\bar\omega}^{2}+3 \right)
 \frac{1}{\bar\rho^2} +   \mathcal{O}(\bar\rho^{-4}) \right) \,,
 \\
 \xi(\bar\rho) \sim &
 - \bar\omega^2 \bar\rho
 -3 \frac{\bar\nu}{\bar{m}}
 +\mathcal{O}(\bar\rho^{-1}) \,.
\end{align}
On the other hand, we impose the incoming wave condition at the horizon
$\bar{\rho}=\bar{\rho}_h$, and normalize the modulus of $z_{\bar\omega}$
to be unity at the horizon so that near the horizon
$z_{\bar\omega}(\bar\rho) =(\bar\rho-\bar\rho_h)^{-i \bar\omega}
\left(1+\mathcal{O}((\bar\rho-\bar\rho_h)) \right)$. With this boundary condition at the horizon, we obtain the boundary
condition of $\xi$ at the horizon, and solve the flow equation towards the
AdS boundary. We then arrive at the spectral function, i.e.,
the imaginary part of the retarded Green function
\be
\text{Im}\, G_R(\bar\omega)= 3 a^2 R^2 T_F  {\bar\nu\over \bar{m}}\,,
\ee
where $\bar{\nu}/\bar{m}$ depends only on $\bar\omega$ and is independent of $R$ and $a$.
For small $\bar\omega$ region, the imaginary part of the retarded
Green function has the form $\text{Im} G_R \sim a_1 \bar\omega +
\mathcal{O}(\bar\omega^2)$, and
according to the exact result \cite{Xiao:2008nr} the factor $a_1$ is
expected to be $-1$.
Numerically we have obtained a value of $-1.00109983$, which means that the
error is of order $10^{-3}$.

After checking that our numerical study goes well, we plot the
spectral function in Fig. \ref{spec_String}.
For comparison, we also
plot the spectral function for the thermal quark which is dual to a
fundamental string in an AdS black hole background, i.e., we need to solve the
flow equation \eq{flow_eq} but with $Q_1= \frac{f(z)}{z^2}$ and
$Q_2 = - \frac{1}{z^2 f(z)}$ for the AdS-Schwarzschild metric
$ds^2_{AdS_5-BH} = \frac{R^2}{z_H^2 z^2} \left(-f(z) dt^2 +
d\mathbf{x}^2 + \frac{dz^2}{f(z)} \right) $ with $f(z)=1-z^4$. We
note that the spectral functions for thermal and accelerated quarks show similar qualitative behavior, that is, with a single resonance
peak.  This implies that we cannot distinguish these two types of
quarks by inspecting their spectral functions qualitatively, but
quantitatively we may be able to do that.

\section{Holographic Mesons in Rindler Space}
\label{sec:Unruh_meson}

\setcounter{footnote}{0}

In QCD the chiral symmetry is dynamically broken by the nonzero quark condensate,
i.e., order parameter, due to the strong interaction. Above a certain critical temperature
the chiral symmetry is naively expected to be restored with zero chiral condensate
because the thermal fluctuations overcome the attractive force between quark and
anti-quark.

Since QCD is strongly coupled, it is difficult to discuss the issue of chiral symmetry
restoration directly. Instead one can study it in the holographic dual, and this was
done in \cite{Mateos:2007vn,Myers:2007we} by considering D7 probe branes in
the AdS black hole background. The transverse scalar
field on the D7 brane is identified as the composite $\bar{q} q$, i.e.,
a meson. Using AdS/CFT dictionary, we can compute the
chiral condensate $\langle \bar{q}q \rangle$ and the mesonic spectral function.

The result of the papers \cite{Mateos:2007vn,Myers:2007we}
show that the meson dissociates/melts above some critical temperature
but the chiral condensate goes to zero only as temperature goes to
infinity because of the current quark mass. In this section, we will study the same issue
for the accelerated meson by embedding the probe D7 branes in the
metric \eq{xiao3}. We find similar feature related to the chiral condensate, meson melting
and spectral functions as for the meson at a finite temperature.
For related field
theory discussions, please see \cite{Unruh:1983ac,Ohsaku:2004rv}, and
also our conclusion section.

In the previous section, we discussed the constantly accelerated string
in its comoving frame, which is dual to an accelerated quark.  Here we would like to
introduce an accelerated meson as the probe D7 brane in the metric \eq{xiao3}.
Because a meson is a composite of quark and anti-quark and an
accelerated quark is described by the endpoint of an open
string~\eq{accF1}, we prepare a probe D7 brane to which this open
string ends perpendicularly. Then, this embedding realizes a meson
which is also accelerated in the same way as its constituent quarks
(and $\bar{q}q$ operator), as seen by the constantly accelerated observer
in the CFT side. Notice that this embedding is different from the
supersymmetric embedding of a D7 brane as discussed in Appendix A.

Thus, a probe D7 brane is static in the comoving frame and does not
depend on the coordinate $\alpha$. The latter comes from the fact that
the energy of a straightly extended string along $r$ coordinate does not
depend on its position in the $\alpha$ coordinate, which is easily seen from
the Nambu-Goto action for the string localized at the $(\alpha,x_2,x_3)$
coordinates, i.e.,
\begin{align}
 S &= - T_{F} \!\int\! d\tau \!\int_{1/a}^{r_{UV}}\!d r
 \frac{R^2}{r^2} = -T_F \!\int\! d\tau
 \left( aR^2 - \frac{R^2}{r_{UV}} \right)\,.
\end{align}
If the location of a D7 brane depends on $\alpha$, then the mass of
a quark being viewed as the endpoint of an open string on the D7 brane depends
on the location $\alpha$ of the quark. This, however,  is not what we would
like to realize.

We also comment that if we embed a probe D7 brane in the supersymmetric way,
the value of chiral condensate is zero even after we move to the comoving frame.
This is because the expectation value of an operator does not depend on the choice of
the coordinate system. We explicitly show this in Appendix A.
Here, we consider a different operator from the one in the supersymmetric embedding,
and thus the vacuum expectation value behaves differently.
In order to discuss the Unruh effects, we take the time coordinate in Rindler space as the time direction and accordingly define $\bar{q}q$ operator, which is then apparently different from the $\bar{q}q$ operator defined in Minkowski spacetime.

Geometrically, we introduce a probe D7 brane in the metric \eq{xiao3} as follows
\begin{center}
\renewcommand\arraystretch{1.5}
   \begin{tabular}{r|cccccccccc}
    & $\tau$ & $\alpha$ & $x_2$ & $x_3$ & $\rho$ & $\theta_1$
    & $\theta_2$ & $\theta_3$ & $w_5$ & $w_6$ \\ \hline
   $D7$ & $\sigma^0$ & $\sigma^1$  & $\sigma^2$ & $\sigma^3$
   & $\sigma^4$ & $\sigma^5$
   & $\sigma^6$ & $\sigma^7$ & $w_5(\rho)$ & $0$ \\
   \end{tabular}
\end{center}
where $\theta_1,\theta_2,\theta_3$ are the coordinates for $S^3$
and $\sigma^\mu$ ($\mu=0,1,\cdots,7$) are the D7 brane worldvolume coordinates.
The shape of the D7 brane is given by $w_5=w_5(\rho)$ and
we have used the rotational symmetry of $(w_5,w_6)$ plane to set $w_6=0$.
In this embedding,
one can easily see that the accelerated open string perpendicularly ends on this D7 brane and then the corresponding meson is also accelerated in Minkowski spacetime, thus the meson is static in the comoving frame.

\begin{figure}[t]
\vspace*{-1.5cm}
\begin{center}
  \includegraphics[width=20em,clip]{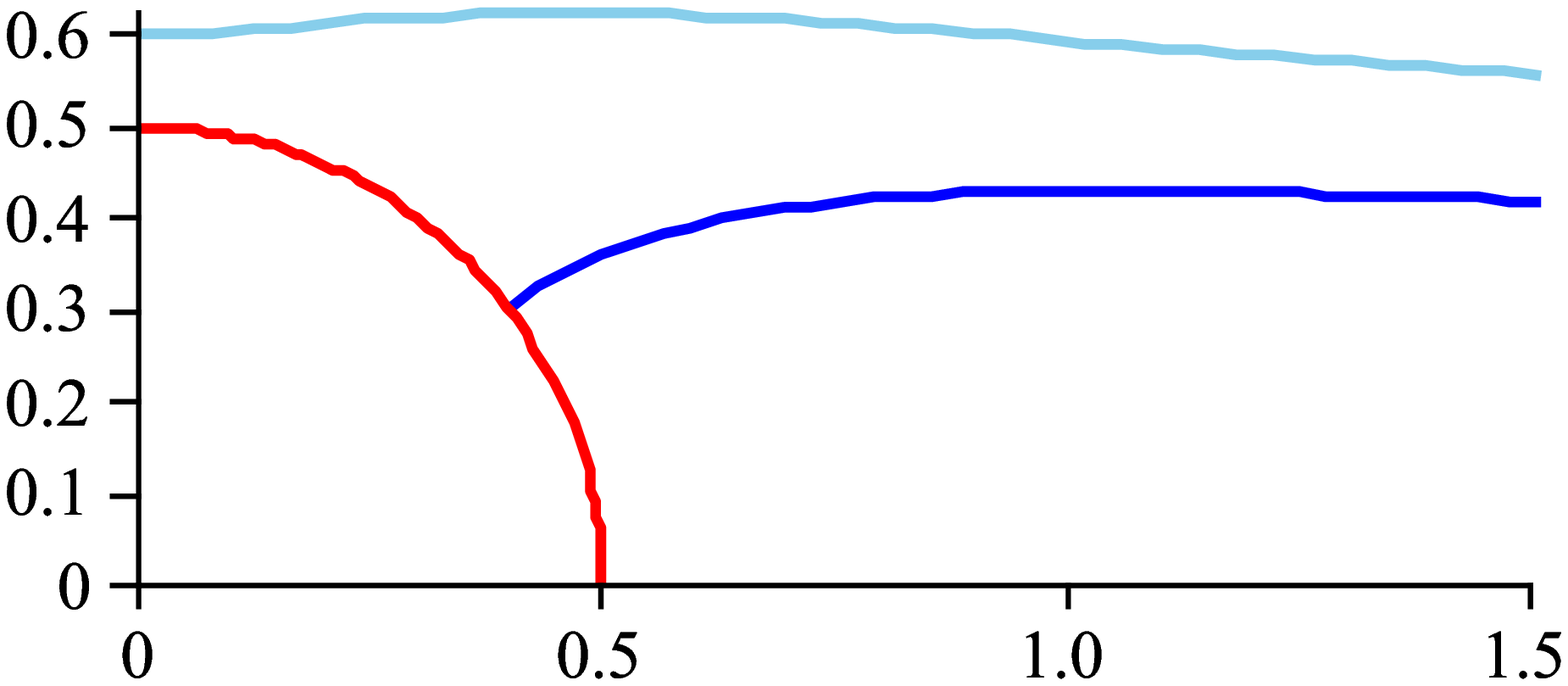}
 \put(-230,170){\makebox(0,0){\large $w_5(\rho)$}}
  \put(-55,165){\makebox(0,0){Minkowski embedding}}
 \put(-60,120){\makebox(0,0){Blackhole embedding}}
 \put(-175,85){\makebox(0,0){horizon}}
\end{center}
\vspace*{-2cm}
 \caption{The typical configurations of a D7 brane in the metric
   (\ref{xiao3}) with $a=R=1$.
   These are the ``Minkowski embedding'' (skyblue line)
   where a D7 brane reaches the center, and the ``Black hole
   embedding'' (blue line) that corresponding to a D7 brane ending on
   the horizon.
 The red line, $w_5^2+\rho^2=1/4$, represents the horizon.}
\label{D7_emb}
\end{figure}

Then the Dirac-Born-Infeld action for the D7 brane becomes
\begin{align}
\label{D7_DBI_leading}
 S_{D7} &= -T_7 \int \! d^8x \sqrt{-\det g}
 \\
 &= -T_7 \int \! d^8x
 e^{-2a\alpha}
 \rho^3
 (1+\frac{a^2R^4}{4w^2})^3(1-\frac{a^2R^4}{4w^2})
 \sqrt{1+w_5'(\rho)^2}
 ,
\end{align}
where $w^2 = \rho^2 + w_5(\rho)^2$ and
$T_7$ is the D7 brane tension.  We then derive the equation of
motion (see Appendix B for details, e.g., \eq{full_eom}), and solve it
with an appropriate boundary condition at $\rho=\infty$. The solution
must be regular everywhere, especially on the horizon. The asymptotic
solution near $\rho=\infty$ is
\begin{align}
 w_5(\rho) &= m \left( 1 + \frac{a^2R^4}{2}\frac{\ln(\rho/a)}{\rho^2}
 +\cdots \right)
 + \frac{\nu}{\rho^2} \left( 1 - \frac{3a^2R^4}{8\rho^2} +\cdots
 \right)
 ,
 \label{asyms}
\end{align}
where $m$ and $\nu$ are constants. The condition that the solution is
regular everywhere determines $\nu$ as a function of $m$. We then
numerically solve the equation of motion and obtain $\nu=\nu(m)$. We
also give a cartoon of the solution in Fig. \ref{D7_emb}.

\subsection{Chiral condensate}
\label{sec:chiral_condensate}

The holographic dual operator for $w_5$ is the composite $\bar{q} q$
and using the AdS/CFT dictionary, we can see that
$m$ in~\eq{asyms} is proportional to
the quark mass and $\nu$ is related to the vacuum expectation value,
i.e. the value of the chiral condensate. In this section, we compute
the value of the chiral condensate and see that it reproduces a proper
behavior at  finite temperature.

Following the standard procedure of holographic renormalization
\cite{Skenderis:2002wp}, we evaluate the D7 brane on-shell DBI action
for the metric \eq{xiao3}, we identify the divergent pieces
by plugging in the asymptotic solution~\eqref{asyms}. The resultant on-shell action becomes
\begin{align}
 S_{D7} &= -T_7 \Omega_3 \!\int\! d^4xd\rho \: e^{-2\alpha a}
 \Big[ \rho^3 +\frac{a^2R^4}{2}\rho -\frac{a^2R^4m^2}{2\rho}
 +\cdots
 \Big]
 \\
 &= -T_7 \Omega_3 \!\int\! d^4x \: e^{-2 \alpha a}
 \Big[ \frac{1}{4}\rho_\infty^4 +\frac{a^2R^4}{4}\rho_\infty^2
 -\frac{a^2R^4m^2}{2}\ln\rho_\infty
 +{\rm (finite)}
 \Big]
 ,
\end{align}
where we regularize the action by introducing the cutoff near the
infinity $\rho=\rho_\infty$ and (finite) stands for the finite pieces in the
limit $\rho_\infty \rightarrow \infty$. We now introduce the counter
term action at $\rho=\rho_\infty$ to cancel the divergences. To
determine the counter term action, we first discuss the symmetry of the
action. It is easily seen that the action is invariant under the scaling
\begin{align}
 \rho \rightarrow k \rho , \hspace{3ex}
 w_5 \rightarrow k w_5 , \hspace{3ex}
 a \rightarrow k' a , \hspace{3ex}
 R \rightarrow \sqrt{k/k'} R , \hspace{3ex}
 \alpha \rightarrow \alpha/k', \hspace{3ex}
 T_7 \rightarrow k'/k^4 \,,
\end{align}
where $k$ and $k'$ are parameters. Under the scaling, $m$ and $\nu$ transform as
$m\rightarrow km$ and $\nu\rightarrow k^3\nu$, and the invariant
combinations under this transformation are $m/\rho$, $aR^2/\rho$ and
$\nu/\rho^3$. Then the counter term  action which keeps this invariance
and cancels the divergence is obtained as follows:
\begin{align}
 S_C &= -T_7\Omega_3 R^4 \!\int\! d^4x \sqrt{-\gamma}
 \Big[
 -\frac{1}{4}
 +( \frac{m^2}{2} - \frac{a^2R^4}{8} )\frac{1}{\rho_\infty^2}
 +a^2R^4m^2\frac{\ln a \rho_\infty}{\rho_\infty^4}
 +c\frac{1}{\rho_\infty^4}
 \Big] ,
\\
 \sqrt{-\gamma} &= e^{-2a\alpha}\frac{w_\infty^4}{R^4}
 (1+\frac{a^2R^4}{4w_\infty^2})^3(1-\frac{a^2R^4}{4w_\infty^2}) ,
 \hspace{3ex}
 w_\infty^2=\rho_\infty^2 + w^2_5(\rho_\infty)
 ,
\end{align}
where $c$ is a constant which yields finite contribution, but cannot be
fixed from the cancellation of divergences.
Due to the above scaling invariance $c$ must take the form of
$(a R^2)^4 f(m/aR^2, \nu/(aR^2)^3)$ where $f$ is a function of these
two invariants to be determined.
In sum, we have
\begin{align}
 S_{D7} + S_C &= -T_7\Omega_3 \!\int\! d^4x e^{-2\alpha a}
 \left[
 -m\nu -\frac{a^4R^8}{16} +\frac{3}{4}m^4 -\frac{a^2R^4}{8}m^2 +c
 +{\cal O}(\rho_\infty^{-2})
 \right] \,.
\end{align}

\begin{figure}[htb]
 \begin{center}
\includegraphics[width=20em,clip]{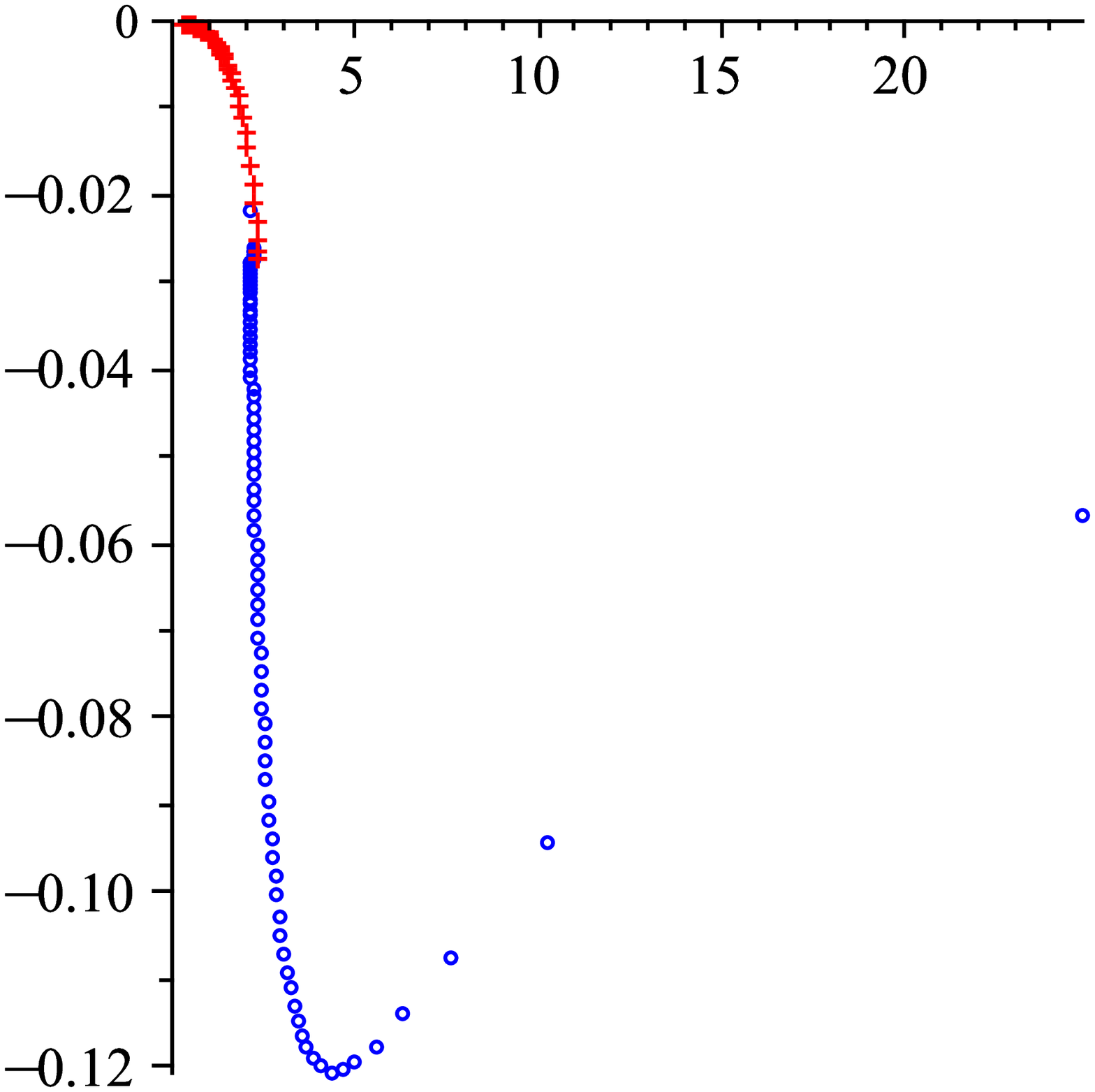}
 \includegraphics[width=20em,clip]{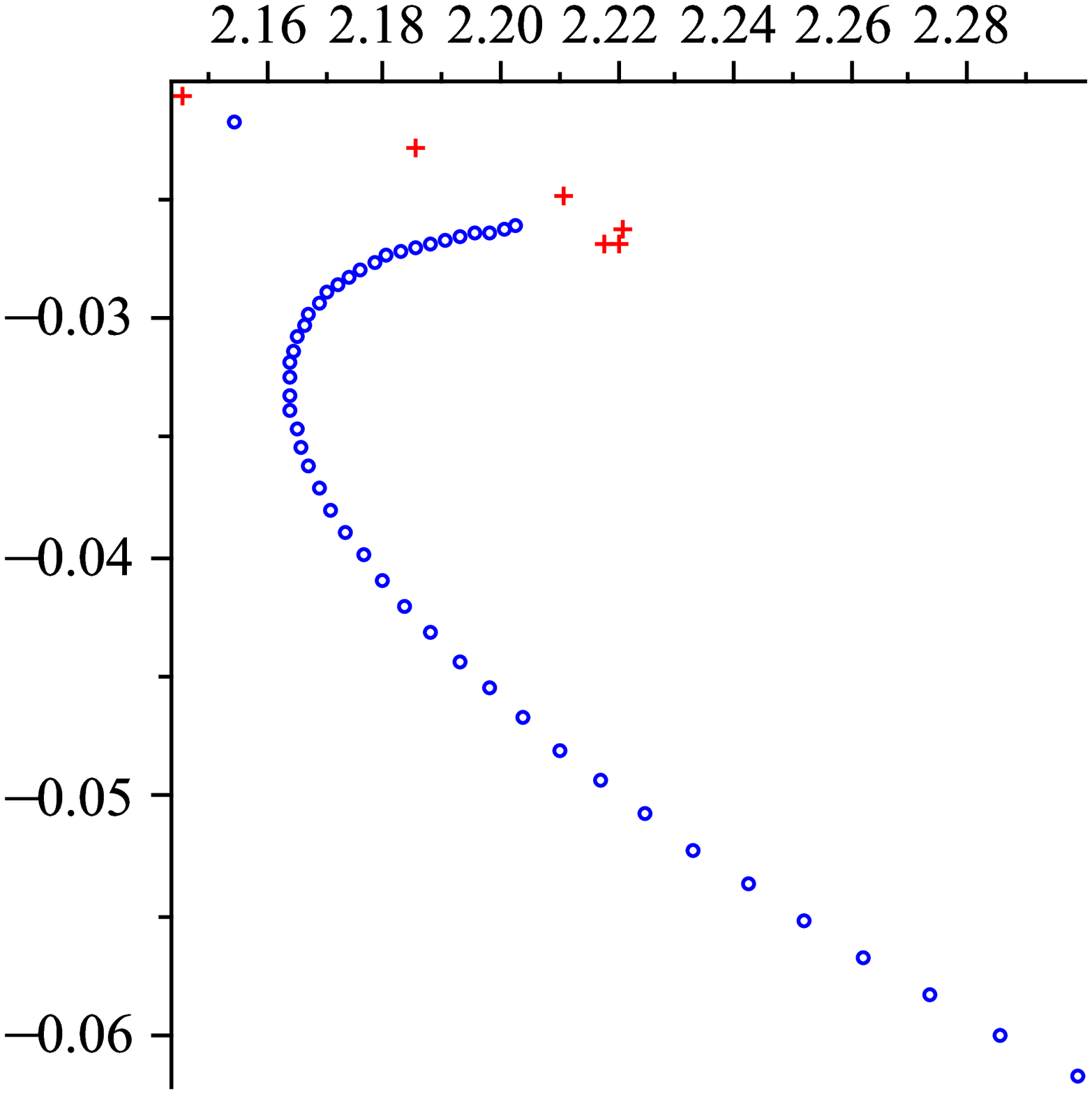}
\put(-350,235){\makebox(0,0){ $1/m$}}
\put(-463,220){\makebox(0,0){ $\langle \bar{q}q \rangle$}}
 \end{center}
 \caption{ {\bf Left:}
 $\langle \bar{q}q \rangle$ against $1/m$ with $R=a=1$.
 {\bf Right:}  the magnification for the transition point of meson melting.
 Red (cross) points for the Minkowski embedding and Blue (circle) ones for the
 Blackhole embedding.  }
 \label{D7-1pt-fcn1}
\end{figure}

Applying the AdS/CFT dictionary, we get the chiral
condensate
\begin{align}
 \langle{ \bar{q}q}\rangle & :=
 \lim_{\rho\rightarrow \infty} \frac{1}{\sqrt{-g^{(0)}}}
 \frac{\delta (S_{D7}+S_C)}{\delta M}
 \nonumber
 \\
 &
 =-\frac{\Omega_3}{(2\pi)^3} N_c \sqrt{\lambda} T^3
\left[
 \left( -3\bar{\nu} + 3\bar{m}^3 + \partial_{\bar{m}}\bar{c} +\mathcal{O}(\rho_\infty^{-2})
 \right) \right]
  \,,
\end{align}
where we have used
\begin{align}
 \frac{\delta S_{D7}}{\delta m}
 =& \frac{\delta S_{D7}}{\delta w_5(\rho)} \frac{\delta w_5(\rho)}{\delta m}
 \nn\\ =&
 \frac{\partial L_{D7}}{\partial w_5'} \bigg|_{\rho=\rho_\infty}
 =-T_7\Omega_3 e^{-2\alpha a} \left(
 -2\nu +\frac{a^2 R^4m}{2} ( 1 -\ln(a \rho_\infty)^2)
 +\mathcal{O}(\rho_\infty^{-2})
 \right) .
\end{align}
The parameters with the bar denote the dimensionless ones,
$\bar{m}=m/aR^2$ etc.
$M$ is the mass of the fundamental matter, $M=T_F m$.
By variating with respect to this mass parameter, we obtain the chiral
condensate.
This expression contains the parameter $c$ which can be fixed from the
following physical reasons. One physical requirement is that in the
supersymmetric limit, $a\rightarrow 0$, the chiral condensate should be
zero. The other requirement is that in the large $m$ limit,
i.e. $m/T\rightarrow\infty$, the chiral condensate again should go to
zero since the temperature effect will be negligible.

The numerical computation shows that $\nu$ goes to zero as $m$
goes to zero, and $\nu/(a^3R^6)\sim -\frac{1}{2}m/(aR^2)\ln m/(aR^2)$
for large $m$ regime.
We thus obtain $\partial_m c = -3m^3 + \frac{3}{2} a^2 R^4 m\ln
\frac{m}{aR^2}$, and
$\langle \bar{q}q\rangle\propto \nu
- \frac{1}{2} a^2 R^4 m\ln \frac{m}{aR^2}$.
We show our numerical results in Fig.~\ref{D7-1pt-fcn1}.
The behavior is qualitatively the same as the AdS black hole case~\cite{Mateos:2007vn},
and thus properly recovers the behavior at finite temperature.
In the left figure of Fig.~\ref{D7-1pt-fcn1}, near the point at
which the two solutions meet, there is a jump of the order parameter.
The right figure is the magnification around this point.
In \cite{Mateos:2007vn}, this jump has been identified as
a first order  phase transition of meson dissociation/melting.
One can see that the behavior of our order parameter is also
quite similar to the one at the finite temperature case in \cite{Mateos:2007vn}.

One comment is in order: the observer is
accelerated along one specific spatial direction and the four
dimensional Lorentz symmetry is broken in CFT which is different from
the Minkowski space in a finite thermal bath. Therefore, the result is
not necessary the same as that in the AdS black hole case.

\subsection{Spectral function of an accelerated meson}

\label{sec:spec_D7}

We now consider fluctuations on the D7 brane and the properties of the mesons.
The fluctuation we consider is the one around the embedding coordinates,
\begin{align}
  w_5 =& z(\rho) + \phi(\tau,\alpha,x_2,\rho) \,,
\end{align}
where to avoid confusion we denote the leading order embedding
function by $z(\rho)$ that used to be $w_5(\rho)$.
We will consider the fluctuations which do not depend on $S^3$
coordinates nor $x_3$ for simplicity. Again, we will use the first order
formalism for the retarded Green function by solving the flow equation.
Numerically, the first order equation is easier to control than
the second order equation of motion as our D7 brane profile in the
previous section, around which we are
considering the retarded Green function for the fluctuations,
is given only numerically.

The holographic retarded Green function for the $\bar{q}q$ operator, which is dual to $M=T_F m$,  is
given by
\begin{align}
\label{def_GR_D7}
  G_R(\bar\omega,\chi) =& T_7 R^{8} \lambda T^2 \Omega_3
 \lim_{\rho\rightarrow \rho_B} \xi(\rho) \,,
\\
  \xi (\rho) =& \frac{\tilde\pi_{(\chi, \bar\omega)}(\rho)}{
  W_{(\chi, \bar\omega)}(\rho)}  \,,
\end{align}
where $W_{(\chi, \bar\omega)}$ and $\tilde\pi_{(\chi, \bar\omega)}$ are
Fourier transform of $\phi$ and its conjugate momentum, respectively. Note that in \eq{def_GR_D7} we have extracted 
all the dimensional parameters in the overall coefficient. 
In Appendix \ref{sec:L_2nd}, we give the details of deriving the flow
equation and the result is
\begin{align}
 \partial_\rho \xi =&
 -\frac{1}{\sqrt{-g} Q_\rho} \xi^2
 +\frac{2V}{ Q_\rho} \xi
 +\sqrt{-g} \left(
 Q_\tau \bar\omega^2 - Q_\alpha \chi
 \right)
 -\frac{\sqrt{-g}}{ Q_\rho} V^2
 +\sqrt{-g} P
 \,.
\end{align}
The explicit forms of $Q_{\tau}$, $Q_{\alpha}$, $Q_{\rho}$, $V$ and $P$
can be found in Appendix B.
Note that we have rescaled all the variables to be dimensionless
as done in (\ref{dimless_emb_var})
and thus the quantities here do not depend on $a$ and $R$.
We put the bar only to $\bar\omega$ to show explicitly this is a 
dimensionless quantity, though all the parameters and functions above
are all dimensionless as well.

We need to impose appropriate boundary conditions to solve the above
flow equation. One way to impose the incoming-wave boundary condition
near the horizon is by taking the asymptotic incoming
solution of $W(\rho)$ near the horizon,
\begin{align}
  W(\rho) \sim & (\rho-\rho_h)^{-i \bar\omega} (1+\mathcal{O}(\rho)) \,,
\end{align}
and directly calculating the $\xi(\rho)$ near the horizon.
This gives the following boundary condition,
\begin{align}
  \xi(\rho) \sim &
 -32 \rho_h^{3} \left( 4\,\rho_h^{2}-1 +4\,i\rho_h^{2}\bar\omega
 \right)
 +\mathcal{O}(\rho-\rho_h)
 \,.
\end{align}
In the same way, we can evaluate the behavior of $\xi$ near the boundary.
The result is
\begin{align}
 \xi(\rho) \sim &
 \frac{1}{2} \left( \chi + \bar\omega^2 + 1 \right)
 \left(1-2 \ln(\rho) \right)
 -2 \frac{C_2}{C_0}
 +\mathcal{O}\left(\frac{\ln(\rho)^2}{\rho^2} \right) \,,
\end{align}
where $C_0$ and $C_2$ correspond to the coefficients of
the asymptotic solution of $W(\rho)$,
\begin{align}
 W(\rho) \sim & C_0 \left( 1+ \mathcal{O}(\rho^{-2}) \right)
 +\frac{C_2}{\rho^2} \left( 1+ \mathcal{O}(\rho^{-2}) \right) \,.
\end{align}
For dimensionful $\tau$, the conjugate frequency is $\omega=a \bar\omega$.
So we have normalized the dimensionless frequency
as $\bar\omega=\omega / 2\pi T$ for the physical frequency $\omega$.

First, we plot the spectral function $-2 \text{Im}G_R$ (more
precisely, just $\text{Im}\, \xi(\rho_B)$) against the dimensionless
frequency $\bar\omega=\frac{\omega}{2\pi T}$.
For the black hole embeddings, the endpoint of the D7 brane is specified by
the value of the $\rho$ coordinate, $\rho_h$.
The position of the horizon is given by
\begin{align}
  w_h^2 = \rho_h^2 + z(\rho_h)^2 = \frac{1}{4} \,,
\end{align}
and then for smaller $\rho_h$ the embeddings are closer to the
 Minkowski ones.
Note that for larger $\rho_h$, the asymptotic value of $z(\rho)$ near
 the boundary, $m$, gets smaller.
Now we can consider $m$ to be the dimensionless quark mass normalized
by the temperature, i.e. $m \sim M_q / T$ \cite{Mateos:2007vn},
 and thus smaller $m$ corresponds to higher temperature.

\begin{figure}[htbp]
 \begin{center}
  \includegraphics[origin=c,angle=270,width=20em,clip]{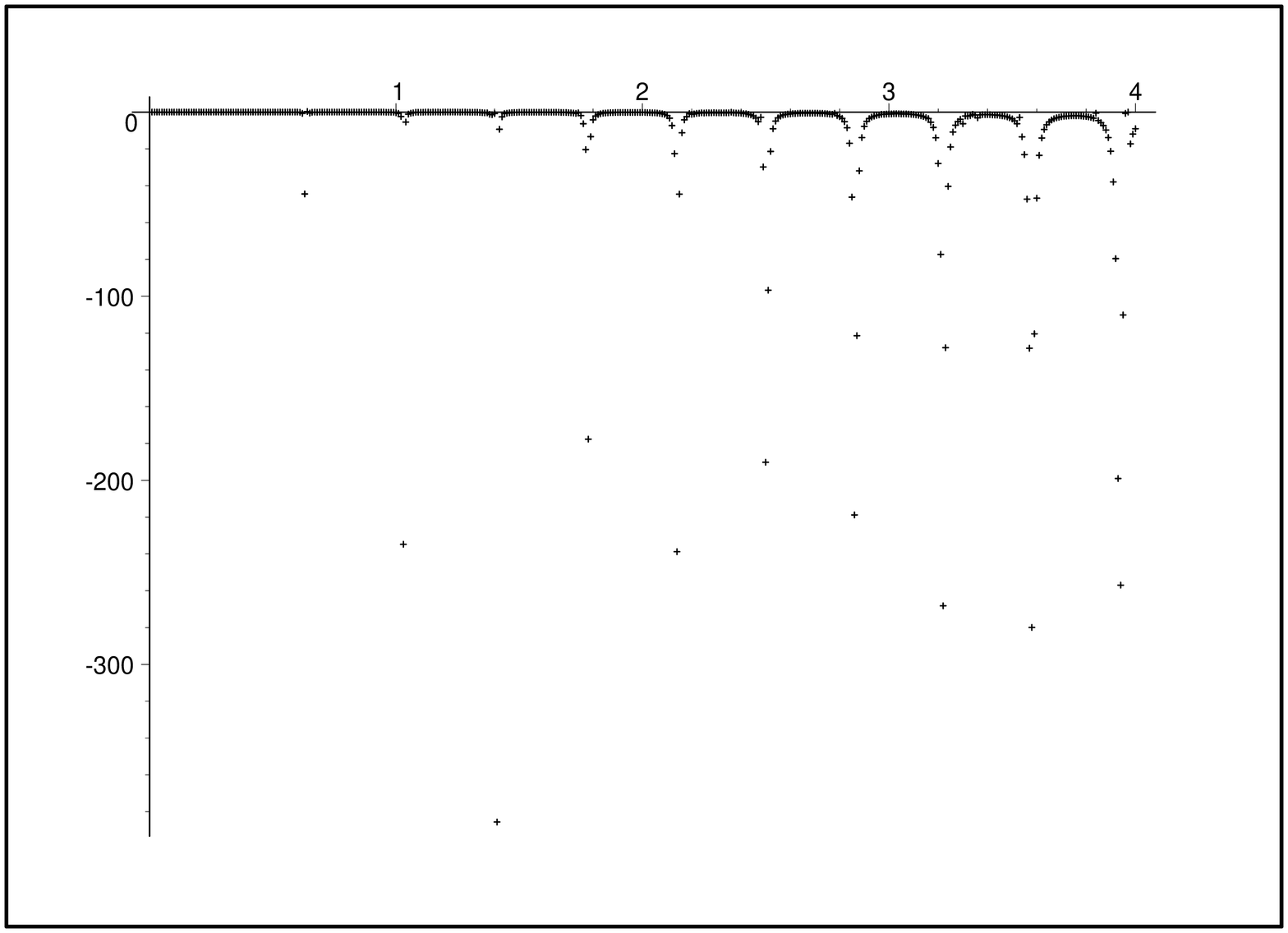}
  \includegraphics[origin=c,angle=270,width=20em,clip]{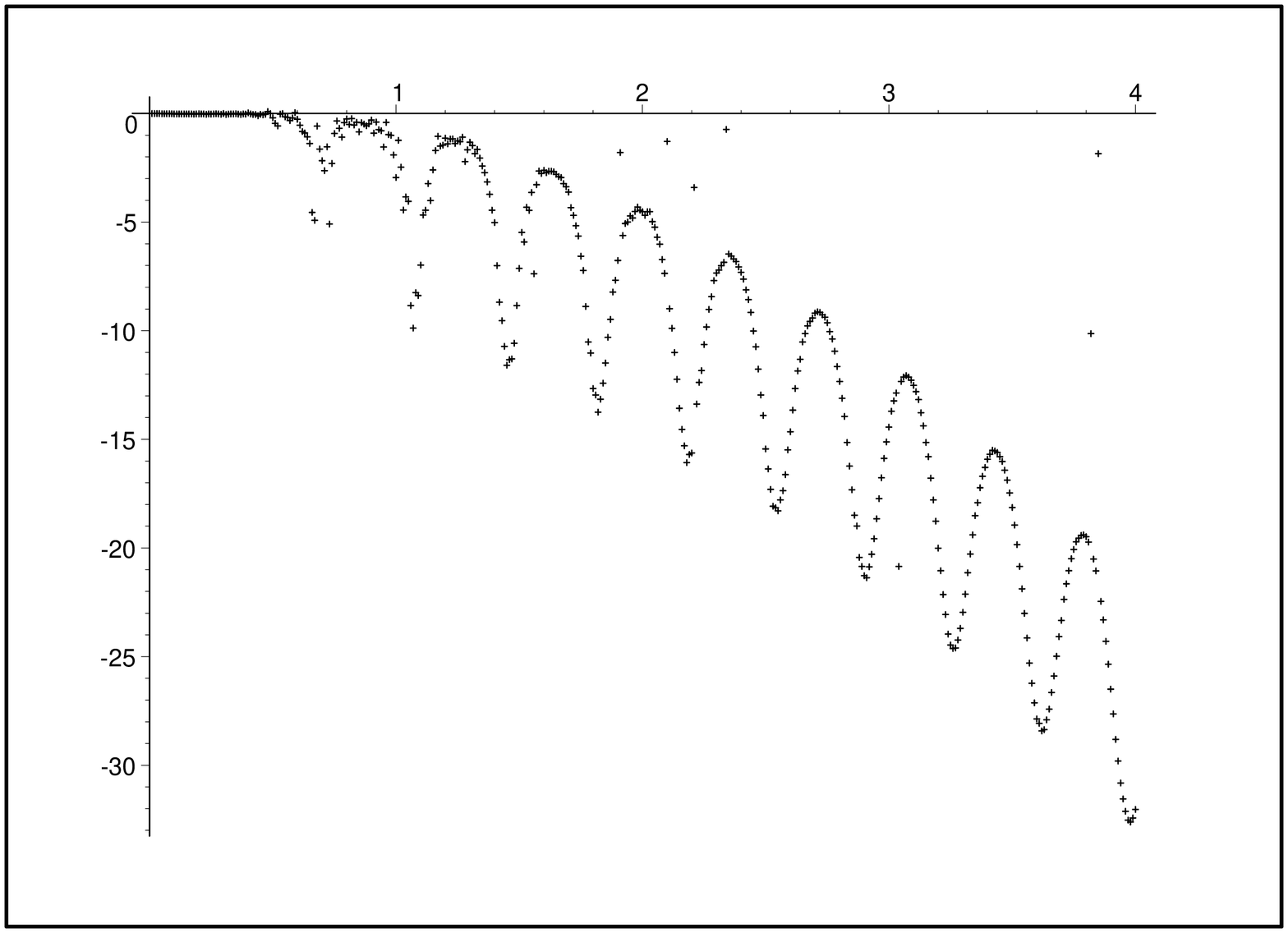}
 \put(-115,195){\makebox(0,0){ $\bar\omega$}}
 \put(-215,42){\makebox(0,0){ $\text{Im}G_R$}}
 \put(-355,195){\makebox(0,0){ $\bar\omega$}}
 \put(-455,42){\makebox(0,0){ $\text{Im}G_R$}}
 \put(-185,55){\makebox(0,0){ $\rho_h=\frac{1}{12}$}}
 \put(-425,55){\makebox(0,0){ $\rho_h=\frac{1}{16}$}}
\\
  \includegraphics[origin=c,angle=270,width=20em,clip]{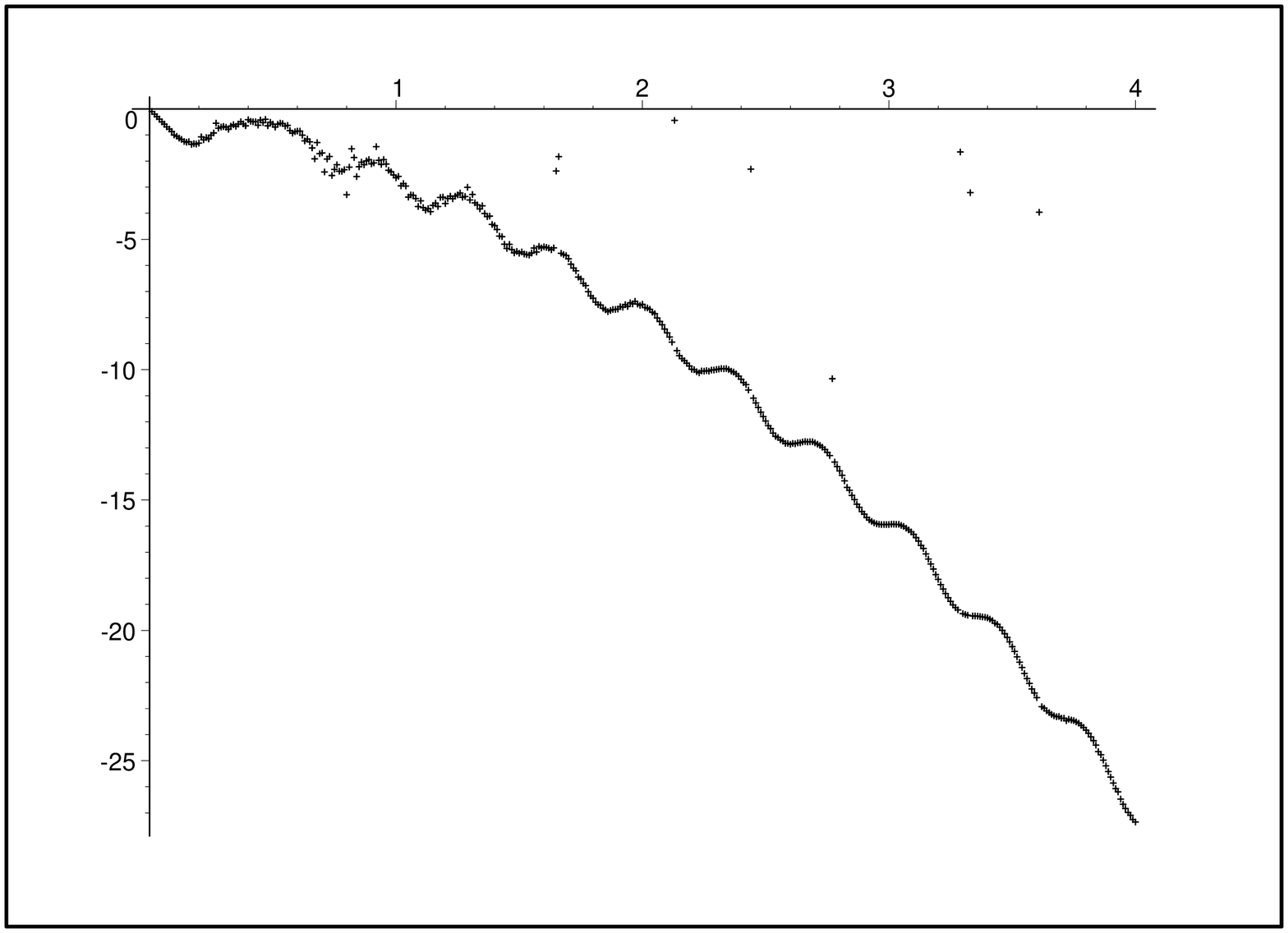}
  \includegraphics[origin=c,angle=270,width=20em,clip]{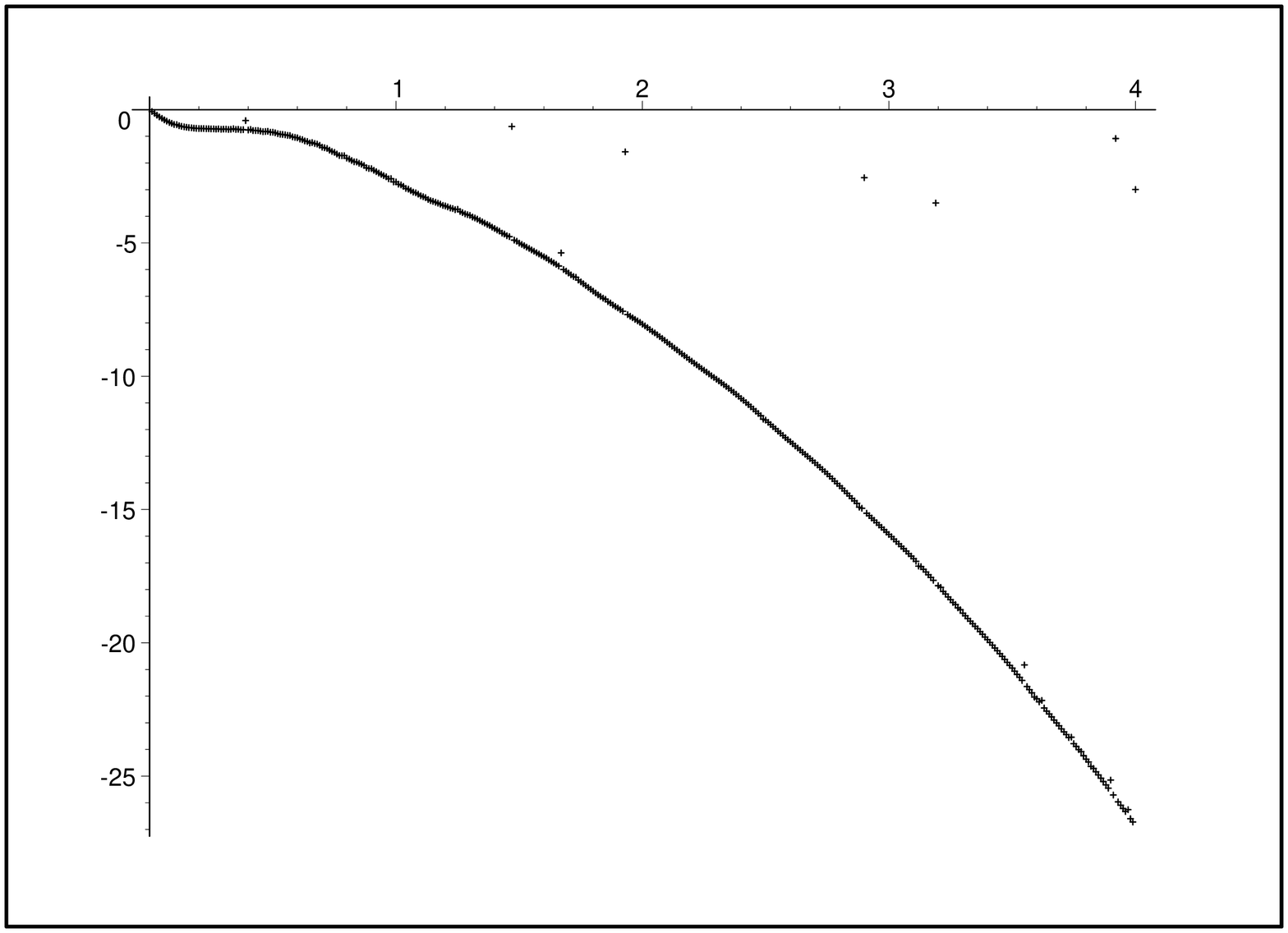}
 \put(-115,195){\makebox(0,0){ $\bar\omega$}}
 \put(-215,42){\makebox(0,0){ $\text{Im}G_R$}}
 \put(-355,195){\makebox(0,0){ $\bar\omega$}}
 \put(-455,42){\makebox(0,0){ $\text{Im}G_R$}}
 \put(-185,55){\makebox(0,0){ $\rho_h=\frac{1}{8}$}}
 \put(-425,55){\makebox(0,0){ $\rho_h=\frac{1}{10}$}}
 \end{center}
 \caption{The spectral functions of the accelerated meson against the
 dimensionless frequency $\bar\omega=\omega/2\pi T$, for the black hole
 embedding.
 $\rho_h=\frac{1}{16},\frac{1}{12},\frac{1}{10},\frac{1}{8}$ as indicated.
It shows that the resonances disappear as the meson melts. }
 \label{D7_spectral_deinsity1}
\end{figure}

In Fig.~\ref{D7_spectral_deinsity1},
 we plot the spectral functions against the dimensionless frequency
$\bar\omega$ for $\rho_h= \frac{1}{16},
\frac{1}{12},\frac{1}{10}$ and $\frac{1}{8}$, with $\chi=0$.
We notice that the spectral function shows sharp peaks for
smaller $\rho_h$ and it gets broaden as $\rho_h$ increases.
This broadening can be understood as the meson spectrum developing
unstable modes for high temperature region due to meson melting.
The qualitative behavior agrees with the one obtained in the
AdS-Schwarzschild analysis \cite{Myers:2007we}.

\begin{figure}[t]
 \begin{center}
 \includegraphics[origin=c,width=15em,clip]{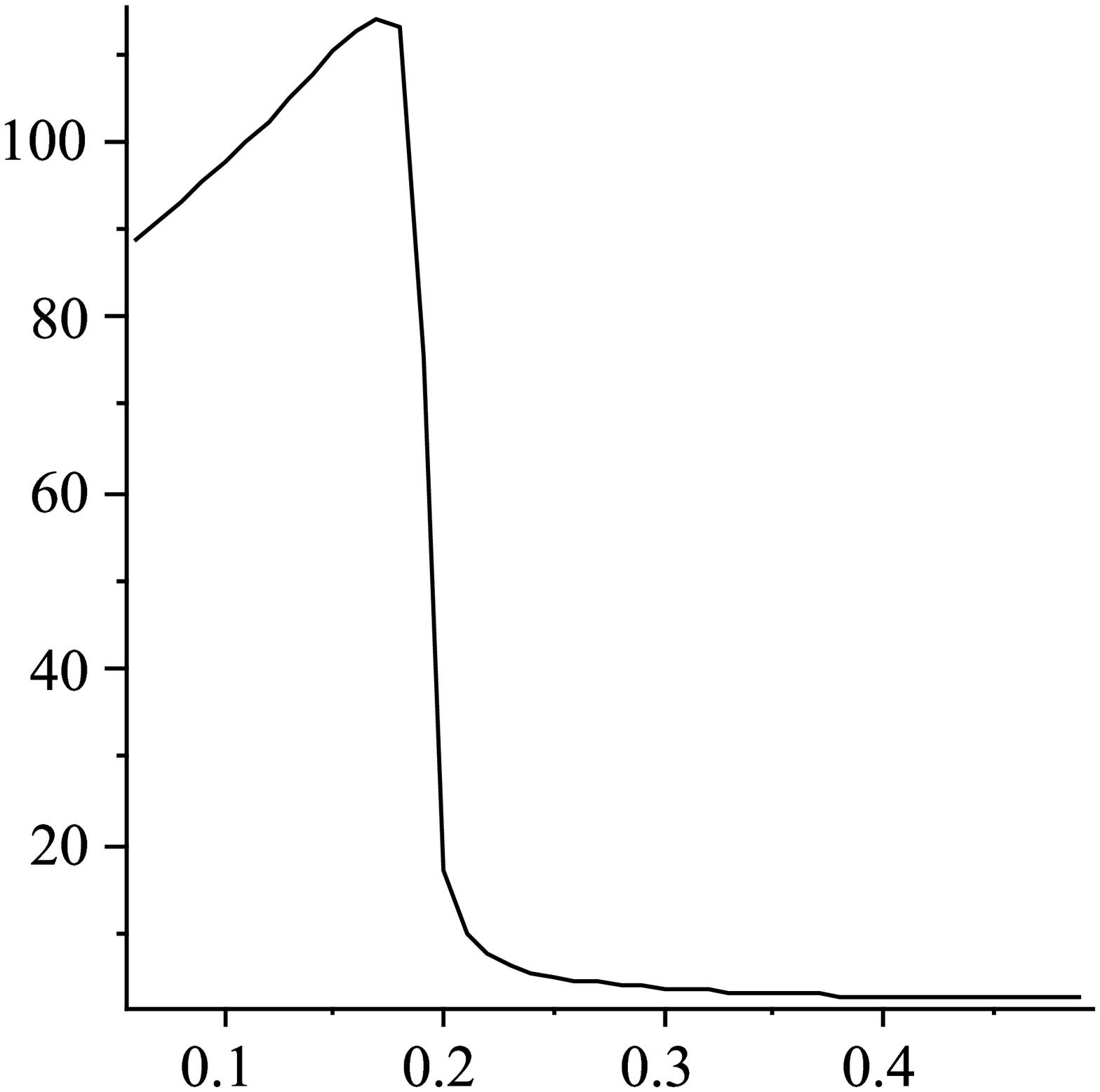}
 \put(-12,7){\makebox(0,0){\Large $\rho_h$}}
 \put(-162,170){\makebox(0,0){\Large $\kappa$}}
 \end{center}
\vspace*{-1em}
 \caption{The strength of the random force $\kappa$ calculated for the
 black hole embedding.
 $\chi$ is set to be $0$.
The point of the jump, $\rho_h \sim 0.19$
 corresponds to $1/m \sim 2.16$ around which the mesons start to melt.
The pre-factor of $\kappa$ is set to be $1$.}
 \label{D7_kappa1}
\end{figure}

We also calculate the strength of the random force through the Kubo
formula, in the black hole embedding, as
\begin{align}
  \kappa =& \lim_{\omega \rightarrow 0} \frac{-2T}{\omega}
 \text{Im} G_R (\omega) \propto
 -  a^3 \, \text{Im} \, a_1 (\rho_h, \chi) \,,
\end{align}
where $a_1$ is a coefficient of $\omega$ in Taylor expansion of
$G_R$ for small $\omega$, i.e. $G_R= a_0 + a_1 \omega + \cdots$.
In Fig.~\ref{D7_kappa1}, we numerically plot $\kappa$ against
$0 \le \rho_h\le 1/2$ for $\chi=0$.
It shows a curious jump around $\rho_h \sim 0.19$.
Actually, $\rho_h=0.18 \sim 0.21$ corresponds to the value of $1/m
=2.164 \sim 2.167$, around which a phase transition of meson melting
occurs in the analysis of the one point function,  as indicated in
Fig.~\ref{D7-1pt-fcn1}.

As noted above, for larger $\rho_h$, $m$ is small and corresponds to a high
temperature regime and for smaller $\rho_h$ it corresponds to a lower
temperature regime, within the black hole embedding.
We give one possible interpretation of this behavior as follows.
For lower temperature, before mesons dissociate, the random force
gets stronger as temperature gets higher.
However, around $\rho_h \sim 0.19$, mesons start to dissociate and then
they behave as a continuum medium rather than rigid particles.
Therefore, the random force from the thermal bath will be dissipated
into these large degrees of freedom incoherently, and the coefficient $\kappa$ takes
a significantly small value.
We may stress that this picture for the behavior of the $\kappa$
is a novel one, and can be used to characterize the meson melting. Also, the sharp drop implies
that meson melting is a first order phase transition.  
 
Again, we can relate this random force to the momentum broadening in the comoving frame by using
Eq. \eq{p-broadening} so that the above sharp transition for $\kappa$ may be observed in the real experiments.
Now the difference from the quark case
is that $\kappa$ is not only a function of $a$, but also of $\rho_h$.
Above the critical $\rho_h$, $\kappa$ becomes constantly small and it
 may suggest that the meson is melting and then the notion of the mean
 square momentum gets lost.
Below the critical $\rho_h$,
we may guess the functional form of $\kappa$ for the black hole embedding\footnote{For sufficiently low temperature, Minkowski embedding should be used
to observe the behavior of $\kappa$.
However we do not pursue the details for this low temperature case,
but rather confine ourselves in the discussion near the transition
point.}  by numerical fitting as
\begin{align}\label{meson-br}
  \kappa =&
T_7 R^8 \lambda T^2 \Omega_3 \frac{2a}{2\pi}
(-a_1(\rho_h))
\sim
a^3 N_c \lambda^2  \left(77.9 + 158.3 \rho_h
+ 375.4 \rho_h^2 + \mathcal{O}(\rho_h^3) \right) \,,
\end{align}
where the fitting function is chosen as a polynomial in $\rho_h$ for
small $\rho_h$ expansion to the quadratic order.
By solving the relation between $\rho_h$ and $m$ numerically,
it is possible to  determine the complicated temperature dependence of $\kappa$, which is quite different from the simple temperature dependence in \eq{xiaokappa} for the quark.
Besides, the $\lambda$ dependence for quark in \eq{xiaokappa} and
meson in \eq{meson-br} are also different, which  may reflect the composite nature of the meson in coupling to the background of quark-gluon plasma.

 However, in the lab frame it is difficult to understand the above behavior of $\kappa$  since the accelerating
neutral particle emits nothing at the leading order, at least in the usual framework of perturbation theory. Therefore, it is not clear what causes the stochastic behavior as characterized by  \eq{meson-br}. It raises the issue that if the complementarity principle of Rindler horizon also works for meson as it works for the quark, at least in the strongly coupled regime. We will leave it for the future study.

\section{Summary and Discussion}

In this paper, we discuss the holographic dual description of the Unruh
effect in large $N_c$ QCD dynamics. A constantly accelerated particle is
identified as
the endpoint of an open string whose value of acceleration depends on the
radial position in AdS space. We prepare a comoving frame and
propose to identify the radial direction in the comoving frame as the
energy scale for the constantly accelerated observer in
CFT\footnote{See \cite{Hirayama:2006jn} where dS${}_4$ space, instead of
 Rindler space, is discussed and a similar identification is done.}.
 We give
supporting evidences by computing holographically the stress energy
tensor, $\bar{q}q$ potential, chiral condensate and the spectral
functions and the diffusion constants of both the accelerated quark and the meson.

In order to discuss the chiral condensate
with fixed Unruh temperature, we introduce D7 brane probe
in such a way that the open strings (dual constituent quarks of meson) on it are static in the Rindler space.
That is, the D7 brane is static in bulk Rindler-like frame, not the Poincare frame.
Since we like to introduce $\bar{q}q$ operator defined by the constantly accelerated observer in CFT, which
is different from $\bar{q}q$ operator defined by the observer in
Minkowski space, our D7 brane probe configuration is different from
the supersymmetric configuration. Our results of the chiral condensate qualitatively show the same behavior as that in the AdS black hole case. We further study its holographic spectral function numerically,
and find that the strength of the random force of Langevin dynamics can
characterize the phase transition of meson melting. Based on our result for the neutral meson, we then raise the issue toward a formulation of complementarity principle for the Rindler horizon.

We briefly comment on the issue about the restoration
of the symmetry such as chiral phase transition for an accelerating
observer. Via Unruh effect, though the local detector accelerated with
the observer does detect real thermal bath of Unruh temperature, some
properties of the vacuum, like spontaneous
symmetry breaking, have been convincingly argued to be independent
of the choice of the observer.
This is expected, since the expectation value (VEV) does not depend on the coordinate system.
If we quantize an operator in Minkowski spacetime (we call this
operator $O_M$) and compute VEV, we obtain the same result whatever
the coordinate system we use. However in order to discuss the Unruh
effect, we quantize an operator in Rindler spacetime (we call this
operator $O_R$), i.e. we use the time coordinate in Rindler spacetime
as the time, the VEV of $O_R$ is different from the VEV of
$O_M$. (Here we also notice that the state with which we take VEV does
not change by the coordinate transformation.) Therefore, if VEV of
$O_M=\bar{q}q_M$ is zero in Minkowski spacetime, it is zero in Rindler
spacetime as well. (See Appendix A and \cite{Unruh:1983ac} for
detailed calculations.) However if we compute $O_R=\bar{q}q_R$, we
obtain nonzero value and this is the Unruh effect. This is what we
have done in our paper. (See also \cite{Ohsaku:2004rv} for example.)

Finally, we like to emphasize that our result is for strong coupling regime obtained from AdS/CFT correspondence, it is not legal to extrapolate it to the weak coupling case by taking $\lambda$ to zero. If we were doing that, we would conclude that the Unruh effect vanishes in the zero coupling limit.   This is not the case, and we should consider the weak coupling case in the framework of usual perturbation theory, where the Unruh effect is argued to be quite robust.

 We still have some open questions. For example,
one may consider more general accelerated string
configurations. Therefore, through  a coordinate transformation to the comoving frame we could obtain
other bulk metrics and here comes a natural question;
do these bulk frames also reproduce the
holographic Unruh effect honestly? Indeed, some earlier works
\cite{Paredes:2008cr} had adopted different bulk metrics based on
other accelerated string configurations.  One may hope that by
requiring the regularity of the derivative expansions of the perturbed
bulk configurations, one could uniquely determine the bulk metric we
choose here. This issue may deserve further study.

\bigskip\bigskip

\subsection*{Acknowledgments}

The authors would like to thank all the members of the string group in
Taiwan. T.H would like to thank the organizers of APCTP Focus Program
Aspects of Holography and Gauge/string duality at APCTP Headquarters,
POSTECH, Pohang Korea and KEK for hospitality,  and thanks Elias
Kiritsis, Koji Hashimoto, Shigeki
Sugimoto, Ho-Ung Yee and Piljin Yi for discussions and comments.
S.K. thanks to KEK for hospitality and the support from the visitor program
and also to the organizers of 2nd Mini Workshop on String Theory
at KEK theory center.
He also thanks
Satoshi Iso, Shin Nakamura and Shiraz Minwalla for
useful comments and discussion.
We also thank the support from the  National
Center for Theoretical Sciences, Taiwan. This work is financially
supported by Taiwan's NSC grant
97-2811-M-003-012 and 97-2112-M-003-003-MY3.

\appendix

\section{Chiral condensate in Rindler space for the supersymmetric embedding}

Since the Unruh effect comes from the fact that the physics is observer
dependent, one may think that we should introduce D7 brane probe in the
same way as the supersymmetric case, and then we only need to go to the
comoving frame.

In this appendix, we will show that this is not the case. In the
supersymmetric case, D7 brane is located at $\theta=0$ and
$\psi=\arcsin m z$ in the AdS$_5\times S^5$ metric
\begin{align}
 ds^2 &= \frac{R^2}{z^2} \left( -dt^2 + dx_1^2 +dx_2^2 + dx_3^2 + dz^2
 \right)
 + R^2 \left( d\psi^2 +\sin^2 \! \psi \hspace{0.1ex} d\theta^2
 + \cos^2 \! \psi  \hspace{0.1ex} d\Omega_3^2  \right)
 .
\end{align}
The induced metric in the comoving frame becomes
\begin{align}
 ds_{D7}^2 &= \frac{R^2}{r^2}\Big[
 \frac{dr^2}{1-a^2r^2} - (1-a^2r^2) d\tau^2 +d\alpha^2
 + e^{-2a\alpha} (dx_2^2 + dx_3^2) \Big]
 \nonumber
 \\
 &\;\;\;
 +R^2 \psi'(z)^2 e^{2a\alpha} \Big[
 dr + ar d\alpha
 \Big]^2
 +R^2\cos^2\psi(z)d\Omega_3^2
\end{align}
Then the D7 brane action becomes
\begin{align}
 S_{D7} &= - T_7 \int \! d^8x \sqrt{-\det g_{ab} }
 \\
 &= - T_7\Omega_3 \int \! d^5x \: \frac{R^8}{r^5} \cos^3\psi(z)
 e^{-2a\alpha} \sqrt{
 1+r^2e^{2a\alpha}\psi'(z)^2 }
 \\
 &= - T_7\Omega_3 \int \! d^4x \! \int_{r_0}^{1/a} \!\! dr\:
 \frac{R^8}{r^5} ( e^{-2a\alpha} -m^2 r^2)
\\
 &= - T_7\Omega_3 R^8\int \! d^4x
 \left( \frac{e^{-2a\alpha}}{4r_0^4} -\frac{m^2}{2r_0^2}
 \right) -
 \left( \frac{e^{-2a\alpha}}{4a^{-4}} -\frac{m^2}{2a^{-2}}
 \right)
 ,
\end{align}
where we introduced the cutoff at $r=r_0$ near the AdS boundary
$r=0$. We now introduce the boundary action at $r=r_0$ as follows
\begin{align}
 S_C &= - T_7\Omega_3 \! \int \! d\tau d\alpha dydz
 \sqrt{-\gamma} \: F[R_\gamma,\psi]
 , \hspace{3ex}
 \\
 \sqrt{-\gamma}& = \frac{R^4}{r_0^4}\sqrt{1-a^2r_0^2}e^{-2\alpha a}
 , \hspace{3ex}
  \psi(r_0e^{a\alpha}) =\arcsin(mr_0e^{a\alpha})
 ,
 \hspace{3ex}
 R_\gamma = -\frac{6a^2r_0^2}{R^2}
 ,
\end{align}
where $\gamma$ is the induced metric at $r=r_0$ and $R_\gamma$ is the Ricci scalar computed from the induced metric. To cancel the divergences, $F$ takes
\begin{align}
 F &= -\frac{R^2}{4} +\frac{R^6}{48}R_\gamma
 +\frac{R^4}{2} \psi(r_0e^{a\alpha})^2
 +c_1\psi(r_0e^{a\alpha})^4
 +c_2R_\gamma \psi(r_0e^{a\alpha})^2 .
\end{align}
The action becomes
\begin{align}
 S_{D7} + S_C &= -T_7\Omega_3 \! \int \! d^4x \Big[
 -\frac{5}{32}a^4R^8e^{-2a\alpha}
 +\frac{R^8a^2}{4}m^2
 +\frac{R^8}{6}m^4e^{2a\alpha}
 +c_1R^4m^4e^{2a\alpha}
 -c_2a^2R^2m^2
 \nonumber\\
 &\;\;\;\;
 +{\cal O}(r_0)
 \Big]\;.
\end{align}
Then, the chiral condensate is given by
\begin{align}
 \langle {\bar{q} q} \rangle & 
\propto \lim_{r_0\rightarrow 0}
 \frac{1}{\sqrt{-\gamma}}
 \frac{\delta (S_{D7}+S_C)}{\delta \psi(r_0e^{a\alpha})}
 =
 -T_7\Omega_3 \frac{r_0^3}{R^4}e^{2a\alpha}
 \Big[
 (\frac{R^8a^2}{2} -2c_2a^2R^2 )m
 +(\frac{2R^8}{3}+4c_1R^4)e^{2a\alpha}m^3
 \Big]
 .
\end{align}
Since we should have vanishing chiral condensate in the $a \rightarrow
0$ limit, we obtain $c_1=-R^4/6$. Also it becomes zero in the large mass
limit, $m\gg a$, we obtain $c_2=R^6/4$ and we finally have
\begin{align}
 \langle \bar{q}q \rangle &= 0
 .
\end{align}
Therefore, the chiral condensate is always zero no matter what $m$ and
$a$ are.

This result is expected since we only change the coordinate system but
do not change the way of the embedding of the D7 brane in the bulk AdS
space.
This means that the meson operator is quantized with respect to
the same time-like Killing direction in both the coordinate systems.
This is consistent with the known result that the vacuum expectation value
is independent of the choice of the coordinate system and the observer
\cite{Unruh:1983ac}.

Note that in the main part of the paper, we consider the chiral
condensate of 
different meson operators by also changing the 
D7 brane embedding, and we therefore
could observe a thermal phase transition
there due to the Unruh effect.


\bigskip

\section{Derivation of flow equation for the spectral function of
  accelerated mesons}
\label{sec:L_2nd}

In this appendix, we will give the detail of the derivation of the
flow equation presented in $\S$ \ref{sec:spec_D7}.

\paragraph{The equations of motion of the fluctuation}

We now embed the D7 brane with fluctuations in
$AdS_5 \times S^5$ with the metric \eq{xiao3}.
We choose the static gauge
\begin{align}
  (\sigma^0, \ldots, \sigma^7)
=& (\tau,\alpha,x_2,x_3,\rho, \Omega_3)
\end{align}
and the embedding is characterized by
\begin{align}
  w_5 =& z(\rho) + \phi(\tau,\alpha,x_2,\rho) \,,
\qquad w_6=0 \,,
\end{align}
where $z(\rho)$ is the leading order contribution specifying the shape
of the D7 determined in Sec. \ref{sec:Unruh_meson} that was
called $w_5(\rho)$.
To avoid the confusion with $w_5$ including the fluctuations,
we use the different name for the same embedding function and
$\phi$ is the fluctuation around this embedding
and is assumed to be independent of $S^3$ coordinates.
We also omitted $x_3$ dependence since we will consider Fourier modes
of this $\phi$ and then by symmetry it is sufficient to consider the
momentum of $x_2$ for modes.
The D7 brane action is given by
\begin{align}
  S_7 =& -T_7  \int d^8 x \sqrt{-\det G_7}
 = T_7 \Omega_3 \int d\tau d\alpha dx_2 dx_3 d\rho
 \left[ \mathcal{L}_0 (z) + \mathcal{L}_2 (z,\phi)
 + \mathcal{O}(\phi^3) \right] \,,
\end{align}
where $\Omega_3$ is the volume of the unit $S^3$ and
$\mathcal{L}_0= \sqrt{-\det G_0}$ is the one in (\ref{D7_DBI_leading}).
The quadratic order action for the fluctuation takes the form
\begin{align}
  \mathcal{L}_2 =&
\frac{1}{2} \sqrt{-\det G_0} \left(
Q_m \partial_m \phi \partial_m \phi
+2 V(\rho) \phi \partial_\rho \phi
+ P(\rho) \phi^2
 \right) \,.
\end{align}
Before giving the explicit form of the coefficient functions $Q_m$,
$V$ and $P$, we introduce the dimensionless variables
\begin{align}
\label{dimless_emb_var}
  (\tau, \alpha, x_2,x_3) = &
 a^{-1} (\bar\tau,\bar\alpha,\bar{x}_2,\bar{x}_3)
\,, \qquad
  (\rho, w_5, w_6) =  a R^2
  (\bar\rho,\bar{w}_5, \bar{w}_6 ) \,.
\end{align}
Thus,
\begin{align}
\label{fluc_action_res}
  S_2 =& -T_7 R^8 \Omega_3
\int d\bar\tau d\bar\alpha d\bar{x}_2
d\bar{x}_3 d\bar\rho
\frac{1}{2} \sqrt{-\det \bar{G}_0} \left(
\bar{Q}_m \partial_{\bar{m}} \phi \partial_{\bar{m}} \bar\phi
+2 \bar{V}(\bar\rho) \bar\phi \partial_{\bar\rho} \bar\phi
+ \bar{P}(\bar\rho) \bar\phi^2
 \right) \,,
\end{align}
where $\bar{m}$ denotes $\bar\tau, \bar\alpha, \bar{x}_2$,
and $\bar{x}_3$.
In $\bar{G}_0(\bar{z})$, all $a$ and $R$ are replaced with $1$
and
\begin{align}
\bar{Q}_{\bar\tau} =&
-\frac{1}{(1+\bar{z}^{\prime 2})w_0^4 \bar{h}_-^2} \,,
\quad
\bar{Q}_{\bar\alpha} =
\frac{1}{(1+\bar{z}^{\prime 2})w_0^4 \bar{h}_+^2} \,,
\quad
\bar{Q}_{\bar\rho} =
\frac{1}{(1+\bar{z}^{\prime 2})^2} \,,
\quad
\bar{Q}_{\bar{x}_2} = \bar{Q}_{\bar{x}_3}
=e^{2 \bar\alpha} \bar{Q}_{\bar\alpha}  \,,
\nn\\
\bar{V}(\bar\rho)=& \frac{1}{2}
{\frac {\bar{z}\,\bar{z}' \left( 1-2 {w_0}^{2} \right)
  }{ \left( 1+\bar{z'}^{2} \right) {w_0}^{6} \bar{h}_+\,\bar{h}_-}} \,,
\nn\\
\bar{P}(\bar\rho)=& \frac {1}{8 {w_0}^{10}{\bar{h}_+}^{2}\bar{h}_-}
\left(-9\bar{z}^{2}+{\bar\rho}^{2}+2{\bar\rho}^{4}-4{\bar\rho}^{2}\bar{z}^{2}
-6\bar{z}^{4}+8\bar{z}^{2}{\bar\rho}^{4}-8{\bar\rho}^{6} +40\,\bar{z}^{4}{\bar\rho}^{2}+24\,\bar{z}^{6}
 \right) \,,
\nn\\
\bar{h}_\pm =& 1 \pm \frac{1}{4 w_0^2} \,,
\qquad
w_0^2 = \bar{\rho}^2 + \bar{z}(\bar\rho)^2 \,,
\end{align}
where the prime ${}'$ denotes $\bar\rho$ derivative.
To make the expressions simpler, we remove the bars
($\bar{\phantom{a}}$) from all the variables
from now on.

From (\ref{fluc_action_res}),
the equations of motion are
\begin{align}
\label{full_eom}
  \sum_{m \neq \rho}
 \partial_m \left(
\sqrt{-G_{0}} {Q_m} \partial_m \phi
\right)
+ \partial_\rho \left[
\sqrt{- G_{0}}
\left( {Q_\rho} \partial_\rho \phi
+V \phi\right)
\right]
-\sqrt{- G_{0}}
\left(
V \partial_\rho \phi + P\phi
\right) =0 \,.
\end{align}
We then introduce the Fourier transform of the field $\phi$ as
\begin{align}
\label{fourier_phi}
  \phi (\tau,\alpha,x_2,\rho)
 =& \int \frac{d\omega d k d \chi }{(2\pi)^2}
 e^{-i\omega\tau + ik x_2} f(\alpha) W(\rho)
\,,
\end{align}
where we have included $x_2$ dependence only by assuming the rotational
invariance.
We introduce a rescaled metric $  g_{mn} = G_{0 mn} \big|_{\alpha=0}$
and the equations of motion becomes ($Q$ denotes $Q_m$ with $m \neq \rho$)
\begin{align}
  \label{eom_rho}
& \partial_\rho \left[
\sqrt{-g} \left(
{Q_\rho} \partial_\rho W +V W
 \right)\right]
-\sqrt{-g} \left(
Q_\tau \omega^2 - Q_\alpha \chi
\right) W
-\sqrt{-g} \left(
V \partial_\rho W + PW
\right) =0 \,,
\\&
\label{eom_alpha}
\partial_\alpha \left(
e^{-2\alpha} \partial_\alpha f(\alpha)
\right)
- \left(
k^2 + \chi e^{-2\alpha}
\right) f (\alpha) =0 \,,
\end{align}
where $\chi$ is a constant that appears in the separation of the
variables.
From these equations of motion, we notice that $W(\rho)$ depends on
$\omega$ and $\chi$ while $f(\alpha)$ on $k$ and $\chi$.
We then sometimes write them as $W_{(\omega,\chi)}(\rho)$ and
$f_{(k,\chi)}(\alpha)$.

As for the $f(\alpha)$ equation of motion,
its general solutions
\begin{align}
\label{f_sol}
  f_{(k,\chi)}(\alpha)=&
c_1 \, e^{\alpha} \,
I_{\nu}(k e^{{\alpha}})
+c_2 \, e^{{\alpha}} \,
K_{\nu}(k e^{{\alpha}})  \,,
\\
\nu =& \sqrt{1- \chi} \,,
\end{align}
can be rearranged into the orthonormal ones
\begin{align}
  \int d\alpha f_{(k,\chi')} (\alpha) f_{(k,\chi)}(\alpha)
  e^{-{2\alpha}}
= \delta_{\chi'+\chi} \,,
\end{align}
for each $k$
with appropriate boundary conditions.
Note that $\chi$ may be discrete or continuous, which depends on the
boundary conditions, but we will mainly consider $\chi=0$ case and then
we just assume that we have chosen an appropriate boundary condition
to obtain an orthonormal set for given $k$.

\paragraph{The retarded Green function}

Suppose that $W_{(\chi,\omega)}$ and $f_{(k,\chi)}(\alpha)$
are solutions to the equations of motion
(\ref{eom_rho}),  and $W_{(\chi,\omega)}$ satisfies the boundary
condition $W_{(\chi,\omega)} (\rho=\rho_B) =
\tilde\phi_0(\omega,\chi)$, where $\rho_B$ stands for the position
of the boundary.

We now consider $\Pi$, the conjugate momentum with respect to $\rho$,
\begin{align}
  \Pi(\tau,\alpha,x_2,\rho) =&
 \frac{\partial \mathcal{L}}{\partial (\partial_\rho \phi)}
=
- T_7 R^8 \Omega_3 \sqrt{- G_0}
\left[ Q_\rho \partial_\rho \phi + V \phi \right] \,.
\end{align}
We may introduce $\Pi= e^{-\frac{2\alpha}{b}} \pi$ where
\begin{align}
\pi =&
- T_7 R^8 \Omega_3 \sqrt{-g}\left[ {Q_\rho}
\partial_\rho \phi + V \phi \right] \,,
\end{align}
and its Fourier mode
\begin{align}
  \pi(\tau,\alpha,x_2,\rho) =
- T_7 R^8 \Omega_3 & \int \frac{d\omega dk \chi}{(2\pi)^2}
  e^{-i\omega\tau + ik x_2} f_{(k,\chi)}(\alpha) \tilde\pi_{(\chi, \omega)}
  (\rho) \,.
\end{align}
In \cite{Iqbal:2008by} Iqbal-Liu has shown that the retarded Green
function on the boundary is obtained via
\begin{align}
  G_R(\omega,k,\chi) = T_7 R^8 \Omega_3
 \lim_{\rho\rightarrow \rho_B}
\frac{\tilde\pi_{(\chi, \omega)}(\rho)}{W_{(\chi, \omega)}(\rho)} \,.
\end{align}
Note that this definition
gives the equivalent Green function to the one by Son-Starinets given
in \cite{Son:2002sd}.

In the papers \cite{Iqbal:2008by,Iqbal:2009fd},
the authors derived a simple differential equation, the flow equation,
which the quantity
\begin{align}
\label{def_xi}
  \xi (\rho) =& \frac{\tilde\pi_{(\chi, \omega)}(\rho)}{
  W_{(\chi, \omega)}(\rho)}
\end{align}
obeys.
By differentiating $\xi$ after $\rho$ and using the equation of motion, we obtain
the  flow equation
\begin{align}
\partial_\rho \xi =&
-\frac{1}{\sqrt{-g} Q_\rho} \xi^2
+\frac{2V}{ Q_\rho} \xi
+\sqrt{-g} \left(
 Q_\tau \omega^2 - Q_\alpha \chi
 \right)
-\frac{\sqrt{-g}}{ Q_\rho} V^2
+\sqrt{-g} P
 \,,
\end{align}
and by solving this equation from the horizon $\rho=\rho_h$ with an
appropriate boundary condition to the boundary $\rho=\rho_B$, we obtain the retarded Green function of the boundary theory,
\begin{align}
  G_R(\omega,\chi) = T_7 R^{8} \Omega_3
 \lim_{\rho\rightarrow \rho_B} \xi(\rho) \,.
\end{align}
This Green function is the one corresponding to the
dimensionless fluctuation $\phi$.
In order to obtain the gauge theory correlator, we need to variate the
action with respect to the mass $M$.
See also $\S$\ref{sec:chiral_condensate}.
Then we get the gauge theory correlator
\begin{align}
  G_R(\omega,\chi) = T_7 R^{8} \lambda T^2 \Omega_3
 \lim_{\rho\rightarrow \rho_B} \xi(\rho) \,.
\end{align}
Here, note that though this result seems to give a simple
temperature dependence of the retarded Green function, the value of
$\xi(\rho_h)$
depends on the choice of some dimensionful parameters such as
$\rho_h$.
So the actual temperature dependence can be more complicated
than seen from this overall scaling.

\bigskip \bigskip


\begin{thebibliography}{99}


\bibitem{Unruh1976}
  W.~G.~Unruh,
  ``Notes on black hole evaporation,''
  Phys.\ Rev.\  D {\bf 14} (1976) 870.

\bibitem{Ford:2005sh}
  G.~W.~Ford and R.~F.~O'Connell,
  ``Is there Unruh radiation?,''
  Phys.\ Lett.\  A {\bf 350} (2006) 17
  [arXiv:quant-ph/0509151]; 
I.~I.~Smolyaninov,
``Photoluminescence from a gold nanotip in an accelerated reference
frame,''
  Phys.\ Lett.\  A {\bf 372} (2008) 7043
 [arXiv:cond-mat/0510743].


\bibitem{Mikhailov:2003er}
  A.~Mikhailov,
  ``Nonlinear waves in AdS/CFT correspondence,'' 
  [arXiv:hep-th/0305196].

\bibitem{Dominguez:2008vd}
  F.~Dominguez, C.~Marquet, A.~H.~Mueller, B.~Wu and B.~W.~Xiao,
  ``Comparing energy loss and $p_{\perp}$-broadening in perturbative QCD with
  strong coupling $\mathcal{N}=4$ SYM theory,''
  Nucl.\ Phys.\  A {\bf 811}, 197 (2008)
  [arXiv:nucl-th/0803.3234].

\bibitem{Susskind:1993ki}
  L.~Susskind,
  ``String theory and the principles of black hole complementarity,''
  Phys.\ Rev.\ Lett.\  {\bf 71}, 2367 (1993)
  [arXiv:hep-th/9307168].


\bibitem{Xiao:2008nr}
  B.~W.~Xiao,
  ``On the exact solution of the accelerating string in AdS(5) space",
 Phys.\ Lett.\  B {\bf 665} (2008) 173
  [arXiv:hep-th/0804.1343].


\bibitem{Deser:1998xb}
  S.~Deser and O.~Levin,
 ``Mapping Hawking into Unruh thermal properties,''
  Phys.\ Rev.\  D {\bf 59}, 064004 (1999)
  [arXiv:hep-th/9809159];
 ``Accelerated detectors and temperature in (anti) de Sitter spaces,''
  Class.\ Quant.\ Grav.\  {\bf 14}, L163 (1997)
  [arXiv:gr-qc/9706018].


\bibitem{Paredes:2008cr}
  A.~Paredes, K.~Peeters and M.~Zamaklar,
  ``Temperature versus acceleration: the Unruh effect for holographic models,''
  JHEP {\bf 0904}, 015 (2009)
  [arXiv:hep-th/0812.0981];
K.~Peeters and M.~Zamaklar,
  ``Dissociation by acceleration,''
  JHEP {\bf 0801}, 038 (2008)
  [arXiv:hep-th/0711.3446].

\bibitem{CasalderreySolana:2006rq}
  J.~Casalderrey-Solana and D.~Teaney,
  ``Heavy quark diffusion in strongly coupled {\cal N} = 4 Yang-Mills,''
  Phys.\ Rev.\  D {\bf 74}, 085012 (2006)
  [arXiv:hep-ph/0605199].

\bibitem{Karch:2002sh}
  A.~Karch and E.~Katz,
  ``Adding flavor to AdS/CFT,''
  JHEP {\bf 0206} (2002) 043
  [arXiv:hep-th/0205236].


\bibitem{Unruh:1983ac}
  W.~G.~Unruh and N.~Weiss,
  ``Acceleration radiation in interacting field theories,''
  Phys.\ Rev.\  D {\bf 29} (1984) 1656;
  C.~T.~Hill,
  ``Can the Hawking effect thaw a broken symmetry?,''
  Phys.\ Lett.\  B {\bf 155} (1985) 343.



\bibitem{Skenderis:2002wp}
  K.~Skenderis,
  ``Lecture notes on holographic renormalization,''
  Class.\ Quant.\ Grav.\  {\bf 19}, 5849 (2002)
  [arXiv:hep-th/0209067].


\bibitem{Maldacena:1998im}
  J.~M.~Maldacena,
  ``Wilson loops in large N field theories,''
  Phys.\ Rev.\ Lett.\  {\bf 80}, 4859 (1998)
  [arXiv:hep-th/9803002];
 S.~J.~Rey and J.~T.~Yee,
  ``Macroscopic strings as heavy quarks in large N gauge theory and  anti-de
  Sitter supergravity,''
  Eur.\ Phys.\ J.\  C {\bf 22}, 379 (2001)
  [arXiv:hep-th/9803001].

\bibitem{WL_finiteT}
S.~J.~Rey, S.~Theisen and J.~T.~Yee,
  ``Wilson-Polyakov loop at finite temperature in large $N$ gauge theory and
  anti-de Sitter supergravity,''
  Nucl.\ Phys.\  B {\bf 527}, 171 (1998)
  [arXiv:hep-th/9803135];
  A.~Brandhuber, N.~Itzhaki, J.~Sonnenschein and S.~Yankielowicz,
  ``Wilson loops in the large N limit at finite temperature,''
  Phys.\ Lett.\  B {\bf 434} (1998) 36
  [arXiv:hep-th/9803137].



\bibitem{Giecold:2009cg}
J.~de Boer, V.~E.~Hubeny, M.~Rangamani and M.~Shigemori,
  ``Brownian motion in AdS/CFT,''
  JHEP {\bf 0907}, 094 (2009)
  [arXiv:hep-th/0812.5112].
  G.~C.~Giecold, E.~Iancu and A.~H.~Mueller,
  ``Stochastic trailing string and Langevin dynamics from AdS/CFT,''
  JHEP {\bf 0907}, 033 (2009)
  [arXiv:hep-th/0903.1840].



\bibitem{quasinormal}
  G.~T.~Horowitz and V.~E.~Hubeny,
  ``Quasinormal modes of AdS black holes and the approach to thermal
  equilibrium,''
  Phys.\ Rev.\  D {\bf 62}, 024027 (2000)
  [arXiv:hep-th/9909056];
 D.~Birmingham, I.~Sachs and S.~N.~Solodukhin,
  ``Conformal field theory interpretation of black hole quasi-normal modes,''
  Phys.\ Rev.\ Lett.\  {\bf 88}, 151301 (2002)
  [arXiv:hep-th/0112055];
 A.~O.~Starinets,
  ``Quasinormal modes of near extremal black branes,''
  Phys.\ Rev.\  D {\bf 66}, 124013 (2002)
  [arXiv:hep-th/0207133]; 
 P.~K.~Kovtun and A.~O.~Starinets,
  ``Quasinormal modes and holography,''
  Phys.\ Rev.\  D {\bf 72}, 086009 (2005)
  [arXiv:hep-th/0506184];
  ``Thermal spectral functions of strongly coupled $N = 4$ supersymmetric
  Yang-Mills theory,''
  Phys.\ Rev.\ Lett.\  {\bf 96}, 131601 (2006)
  [arXiv:hep-th/0602059].


\bibitem{Iqbal:2008by}
  N.~Iqbal and H.~Liu,
  ``Universality of the hydrodynamic limit in AdS/CFT and the membrane
  paradigm,''
  Phys.\ Rev.\  D {\bf 79}, 025023 (2009)
  [arXiv:hep-th/0809.3808].


\bibitem{Son:2002sd}
  D.~T.~Son and A.~O.~Starinets,
  ``Minkowski-space correlators in AdS/CFT correspondence: Recipe and
  applications,''
  JHEP {\bf 0209} (2002) 042
  [arXiv:hep-th/0205051].


\bibitem{Mateos:2007vn}
  D.~Mateos, R.~C.~Myers and R.~M.~Thomson,
  ``Thermodynamics of the brane,''
  JHEP {\bf 0705} (2007) 067
  [arXiv:hep-th/0701132].

\bibitem{Myers:2007we}
  R.~C.~Myers, A.~O.~Starinets and R.~M.~Thomson,
  ``Holographic spectral functions and diffusion constants for fundamental
  matter,''
  JHEP {\bf 0711} (2007) 091
  [arXiv:hep-th/0706.0162].

\bibitem{Ohsaku:2004rv}
  T.~Ohsaku,
  Phys.\ Lett.\  B {\bf 599} (2004) 102
  [arXiv:hep-th/0407067];
  D.~Ebert and V.~C.~Zhukovsky,
  ``Restoration of dynamically broken chiral and color symmetries for an
  accelerated observer,''
  Phys.\ Lett.\  B {\bf 645} (2007) 267
  [arXiv:hep-th/0612009].


\bibitem{Hirayama:2006jn}
 T.~Hirayama,
 ``A holographic dual of CFT with flavor on de Sitter space,''
 JHEP {\bf 0606} (2006) 013
 [arXiv:hep-th/0602258].





\bibitem{Iqbal:2009fd}
  N.~Iqbal and H.~Liu,
  ``Real-time response in AdS/CFT with application to spinors,''
  Fortsch.\ Phys.\  {\bf 57}, 367 (2009)
  [arXiv:hep-th/0903.2596].






\end{thebibliography}
\end{document}